\documentclass[12pt]{book}
\usepackage{sussp}
\usepackage{graphicx}
\usepackage{latexsym}
\usepackage{makeidx}
\usepackage{bm}
\begin{document}

\headings{Lattice QCD}
{Lattice QCD}
{Christine Davies}
{University of Glasgow, UK}

\section{Introduction}%
Lattice QCD \index{lattice QCD}was invented, way ahead of its time, in
1974. It really became a useful technique in the 1990s when a huge
amount of progress was made in the understanding and reduction of
systematic errors. Now, we are poised to start a second lattice
revolution with the onset of Teraflop supercomputing around the world
and further improvements in methodology.  This will enable
calculations using lattice QCD to reach errors of a few percent, over
the next five years. At this level, lattice results, where they exist,
will be the theoretical calculations of choice for the experimental
community.

It seems, then, a good time to review the fundamentals of lattice QCD,
for an audience of experimental particle physicists. As `consumers' of
lattice calculations, it is important to be aware of how these
calculations are done so that a critical assessment of different
results can be made.  I have tried to keep technical details to a
minimum in what follows but it is necessary to understand some of
them, to appreciate the significance and the limitations of the
lattice results that you might want to use. For a more detailed
discussion see, for example~(Gupta, 1998) or (Di Pierro, 2001).  This
school is largely concerned with CP violation and heavy quark physics,
so in Section 4 I concentrate on lattice results relevant to these
areas.

\section{Lattice QCD formalism and methods}

\subsection{The path integral approach}

\begin{figure}
\centerline{\includegraphics[height=5.0cm]{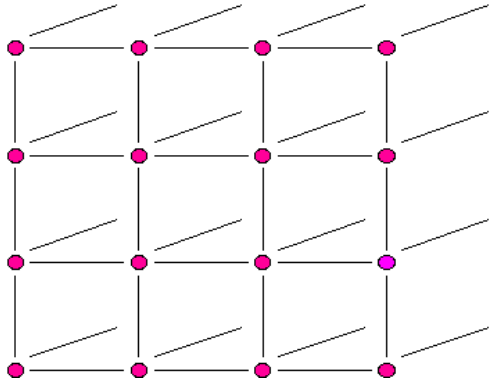}}
\caption{A 2-dimensional rendition of a 3-dimensional cubic lattice. Lattice QCD calculations 
use a 4-dimensional grid.}
\label{lattice}
\end{figure}

Lattice QCD is just QCD, no more and no less. We take the theory,
express it in Feynman Path Integral language, and calculate the
integral as well as we can. We would like to be able to do this in the
continuous space-time of the real world, but this is not
possible. Instead, we must break space-time up into a 4-d grid of
points, i.e. a lattice (\Figure{\ref{lattice}}), and evaluate the
Feynman Path Integral by Monte Carlo methods on a computer. It turns
out to be a calculation that requires a huge amount of computing power
and tests the fastest supercomputers that we have.

In the Feynman Path Integral approach, we first express the quantity
that we want to calculate as the matrix element in the vacuum of an
operator, $\bm{\mathcal{O}}$, which will be a product of quark and gluon fields so
that, for example, 
$\bm{\mathcal{O}} = (\overline{\psi}\psi)_y (\overline{\psi}\psi)_x$.
creates a hadron at a point $x$ and destroys it
at a point $y$. 
We will discuss later other forms that $\bm{\mathcal{O}}$ might take to calculate
useful quantities.  Then:
\begin{equation}
\langle 0|\bm{\mathcal{O}}|0 \rangle =
\Frac {\int{[d \psi]\,[d \bar{\psi}]\,[d A_{\mu}] \,\bm{\mathcal{O}}[\psi,\overline{\psi},A] e^{-S}}}
{\int{[d \psi]\,[d \bar{\psi}]\,[d A_{\mu}] e^{-S}}}
\label{fpi}
\end{equation}
where $S$ is the action, the integral of the Lagrangian: 
\begin{equation}
S = \int\!{ {\cal{L}}\, d^4x}.
\end{equation}
We are using Euclidean space here (imaginary time) so that the integrand
doesn't contain the oscillatory $e^{iS}$, but the more easily 
integrated $e^{-S}$. The integral of Equation~\ref{fpi} can then be 
evaluated numerically if we can convert it to a finite-dimensional 
problem. 

Currently the integral runs over all values of the quark and gluon
fields $\psi$ and $A$ at every point in space-time.  We need to make
the number of space-time points (and therefore field variables) finite
and we do this by taking a 4-d box of space-time and discretising it
into a cubic grid, or lattice. It is then a relatively simple matter
to transcribe the continuous theory onto the lattice, and we use the
standard methods used for discretising e.g. differential equations for
numerical solution.  Continuous space-time $(x,t)$ becomes a grid of
labelled points, $(x_i,t_i)$ or $(n_ia,n_ta)$ where $a$ is the spacing
between the points, called the lattice spacing. The fields are then
associated only with the sites, $\psi(x,t) \rightarrow
\psi(n_i,n_t)$. The action must also be discretised, but this is also
straightforward. The Lagrangian typically contains fields and
derivatives of fields. The fields are replaced with fields at the
lattice sites and the derivatives replaced with finite differences of
these fields. The integral over space-time of the Lagrangian becomes a
sum over all lattice sites: ($\int{d^4x} \rightarrow \sum_n a^4$).
There are inevitably discretisation errors associated with this
procedure (just as there are for differential equations) because the
lattice Lagrangian only matches the continuum Lagrangian at $a$ =
0. At non-zero $a$ there are effectively additional unwanted terms in
the lattice Lagrangian that are proportional to powers of $a$. We will
discuss this further later.  Another view of the lattice is that it
provides an ultra-violet cut-off on the theory in momentum space,
since no momenta larger than $\pi/a$ make sense (the wavelength is
then smaller than $a$). In this way it is an alternative
regularisation of QCD.

As an illustration of the simplicity of the discretisation procedure, 
let us consider a scalar field theory with 
Lagrangian
\begin{equation}
{\cal{L}} = \frac{1}{2}(\partial_{\mu} \phi)^2
+ \frac{1}{2}m^2\phi^2 + \lambda \phi^4.
\end{equation}
The lattice action, $S$ is then 
\begin{equation}
S = \sum_n a^4 \left( \frac{1}{2} \sum_{\mu = 1}^4
\left[\frac{\phi(n+1_{\mu})-\phi(n-1_{\mu})}{2a}\right]^2 
+ \frac{1}{2}m^2\phi^2(n) + \lambda \phi^4(n) \right) . 
\end{equation}
The point $n+1_{\mu}$ is one lattice point up from the point $n$ in
the $\mu$ direction.  We are always free to rescale parameters and
fields and we do this on the lattice, rescaling by powers of the
lattice spacing, so that the parameters and fields we work with are
dimensionless.  Everything is then said to be in `lattice units'.  In
the scalar theory above we rescale to primed quantities where $\phi' =
\phi a,\; m' = ma,\; \lambda' = \lambda$. Then
\begin{equation}
S = \sum_n  \left( {\phi'}^2(n)\left[2 + \frac{1}{2}{m'}^2\right] + \lambda'{\phi'}^4  
- \frac{1}{4} \sum_{\mu} \phi'(n+1_{\mu})\phi'(n-1_{\mu})  \right) .
\label{sclr}
\end{equation} 
The rescaling has the effect of removing the lattice spacing explicitly 
from the action. A lattice calculation is done then without input of 
any value for the lattice spacing, or even knowing what it is. We will 
discuss later converting results back from lattice units to physical units,
so that we can compare results to the real world. Equation~\ref{sclr} has 
in addition been rearranged to collect similar lattice terms together,
using $\sum_n$ to move the space-time indices. 
It now looks very like a spin model, revealing a deep connection between 
lattice field theory and the statistical mechanics of spin systems.

\subsection{Lattice gauge theories for gluons}

\index{gluons in lattice QCD}To discretise gauge theories such as QCD
onto a lattice requires a little additional thought because of the
paramount importance of local gauge invariance. The r\^{o}le of the
gluon (gauge) field in QCD is to transport colour from one place to
another so that we can rotate our colour basis locally. It should then
seem natural for the gluon fields to `live' on the links connecting
lattice points, if the quark fields `live' on the sites.

\begin{figure}
\centerline{\includegraphics[width=11.0cm,clip,trim= 0 35 0
0]{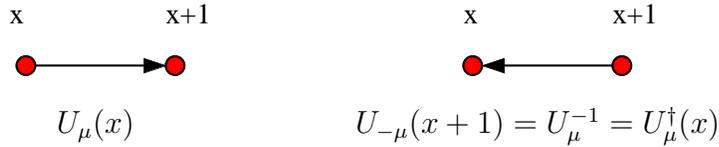}}\vspace{1ex}
{\rule{32mm}{0mm}$U_\mu(x)$\rule{28mm}{0mm}
$U_{-\mu}(x+1)=U_\mu^{-1}=U_\mu^\dagger(x)$}\\
\caption{The gluon field on the lattice.}
\label{gluon}
\end{figure}

The gluon field is also expressed somewhat differently on the lattice
to the continuum. The continuum $A_{\mu}$ is an 8-dimensional vector,
understood as a product of coefficients $A^b_{\mu}$ times the 8
matrices, $T_b$, which are generators of the SU(3) gauge group for
QCD. On the lattice it is more useful to take the gluon field on each
link to be a member of the gauge group itself i.e. a special
(determinant = 1) unitary $3\times3$ matrix. The lattice gluon field
is denoted $U_{\mu}(n_i,n_t)$, where $\mu$ denotes the direction of
the link, $n_i,n_t$ refer to the lattice point at the beginning of the
link, and the color indices are suppressed.  We will often just revert
to continuum notation for space-time, as in $U_{\mu}(x)$.  The lattice
and continuum fields are then related exponentially,
\begin{equation}
U_{\mu} = e^{iagA_{\mu}} 
\label{ulattcont}
\end{equation}
where the $a$ in the exponent makes it dimensionless, and we include the 
coupling, $g$, for convenience.
If $U_{\mu}(x)$ is the gluon field connecting the points $x$ and 
$x+1_{\mu}$
(see \Figure{\ref{gluon}}), 
then the gluon field connecting these same points but in the 
downwards direction must be the inverse of this matrix, $U_{\mu}^{-1}(x)$.
Since the $U$ fields are unitary matrices, satisfying $U^{\dag}U=1$, 
this is then $U^{\dag}_{\mu}(x)$. 

This form for the gluon field makes it possible to maintain exact
local gauge invariance on a lattice. \index{gauge invariance, lattice
QCD}To apply a gauge transformation to a set of gluon fields we must
specify an SU(3) gauge transformation matrix at each point. Call this
$G(x)$. Then the gluon field $U_{\mu}(x)$ simply gauge transforms by
the (matrix) multiplication of the appropriate $G$ at both ends of its
link. The quark field (a 3-dimensional colour vector) transforms by
multiplication by $G$ at its site.
\begin{eqnarray}
U_{\mu}^{(g)}(x) &=& G(x)U_{\mu}(x)G^{\dag}(x+1_{\mu}) \nonumber \\
\psi^{(g)}(x) &=& G(x)\psi(x) \nonumber \\
\overline{\psi}^{(g)}(x) &=& \overline{\psi}(x)G^{\dag}(x)
\label{gts}.
\end{eqnarray}
To understand how this relates to continuum gauge transformations
try the exercise of setting $G(x)$ to a simple U(1) transformation,
$e^{i\alpha(x)}$, and show that Equation~\ref{gts} is equivalent to the 
QED-like gauge transformation in the continuum, $A^g_{\mu} =
A_{\mu} - \partial_{\mu}\alpha$.  

\begin{figure}[!h]
\centerline{\includegraphics[height=5.0cm]{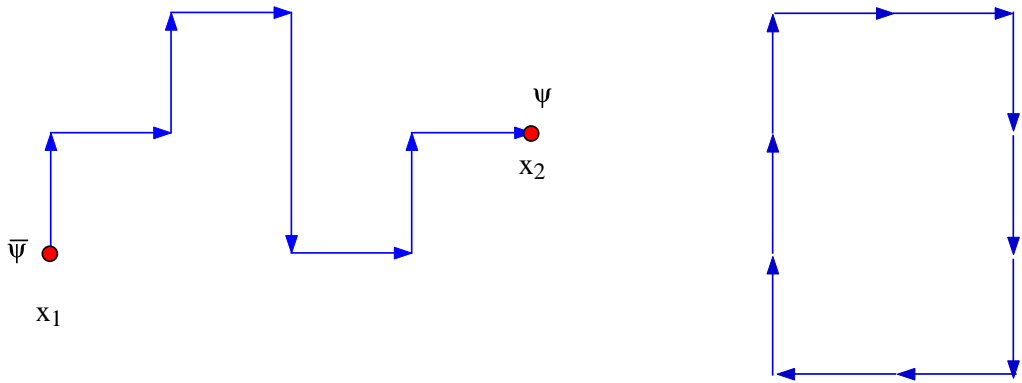}}
\caption{A string of gluon fields connecting quark and antiquark
fields (left) and a closed loop of gluon fields (right).}
\label{string}
\end{figure}

Gauge-invariant objects can easily be made on the lattice out of
closed loops of gluon fields or strings of gluon fields
(\Figure{\ref{string}}) with a quark field at one end and an antiquark
field at the other, e.g.\
$\overline{\psi}(x_1)U_{\mu}(x_1)U_{\nu}(x_1+1_{\mu})\ldots
U_{\epsilon}(x_2-1_{\epsilon})\psi(x_2)$.  Under a gauge transformation
the $G$ matrix at the beginning of one link `eats' the $G^{\dag}$ at
the end of the previous link, since $G^{\dag}G=1$. The $G$ matrices at
$x_1$ and $x_2$ are `eaten' by those transforming the quark and
anti-quark fields, if we sum over quark and antiquark colors. The same
thing happens for any closed loop of $U$s, provided that we take a
trace over color indices. Then the $G$ at the beginning of the loop
and the $G^{\dag}$ at the end of the loop, the same point for a closed
loop, can `eat' each other. (Try this as an exercise, remembering that
$U$ fields going in the downward direction are $U^{\dag}$s and, from
Equation~\ref{gts}, $U_{\mu}^{\dag,(g)}(x) = G(x+1_{\mu})
U_{\mu}^{\dag}(x)G^{\dag}(x)$.)

The purely gluonic piece of the continuum QCD action is
\begin{equation}
S_{\rm cont} = \int d^4x \frac{1}{4g^2}  \Tr F_{\mu\nu}F^{\mu\nu}
\label{sgcont}
\end{equation}
and the simplest lattice discretisation of this is the so-called 
Wilson plaquette action:
\begin{equation}
S_{\rm latt} = \beta \sum_p 
\left(1 - \frac{1}{3} \Re \{\Tr U_p\}\right);\quad
\beta = \frac{6}{g^2}.
\label{sglatt}
\end{equation}
$U_p$ is the closed $1\times 1$ loop called the plaquette, an SU(3) 
matrix 
\begin{figure}[!h]
\centerline{\includegraphics[height=2.0cm,clip,trim=0 7 0 -2]{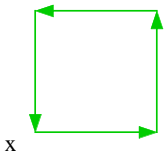}}
\caption{A plaquette on the lattice.}
\label{plaq}
\end{figure}
formed by multiplying  4 gluon links together in a sequence. 
For the plaquette with corner $x$ in the $i,j$ plane we have (\Figure{\ref{plaq}}):
\begin{equation}
U_p(x) = U_i(x)U_j(x+1_i)U^{\dag}_i(x+1_j)U^{\dag}_j(x)
\label{Uplaq}
\end{equation}
Tr in $S_{\rm latt}$ denotes taking the trace of $U_p$ i.e. the sum of the
3 diagonal elements. 
$S_{\rm latt}$ sums over all plaquettes of all orientations on the
lattice.  $\beta$ is a more convenient version for the lattice of the
QCD bare coupling constant, $g^2$. This is the single input parameter
for a QCD calculation (whether on the lattice or not) involving only
gluon fields. Notice that the lattice spacing is not explicit
anywhere, and we do not know its value until {\it after} the
calculation. The value of the lattice spacing depends on the bare
coupling constant. Typical values of $\beta$ for current lattice
calculations using the Wilson plaquette action are $\beta \approx
6$. This corresponds to $a \approx$ 0.1fm. Smaller values of $\beta$
give coarser lattices, larger ones, finer lattices.  Other improved
discretisations of the gluon action are also used. In these the bare
coupling constant appears in a different way and so comparison of the
bare coupling constant between different gluon lattice actions is
meaningless.  The only comparison which makes sense is that of the
resulting values for the lattice spacing.  That $S_{\rm latt}$ of
Equation~\ref{sglatt} is a discretisation of $S_{\rm cont}$ is not
obvious, and we will not demonstrate it here. It should be clear,
however, from Equations~\ref{ulattcont} and~\ref{Uplaq} that
$S_{\rm latt}$ does contain terms of the form $\partial_{\mu}A_{\nu}$.

$S_{\rm latt}$ is gauge-invariant, as will be clear from our earlier
discussion.  Thus lattice QCD calculations do {\it not} require gauge
fixing or any discussion of different gauges or ghost terms.  We
simply calculate the appropriate Feynman Path Integral using
$S_{\rm latt}$.  Since we are only describing calculations for gluons at
this stage, $\bm{\mathcal{O}}$ will be some gauge-invariant product of $U$ fields,
for example the closed loop of Figure~\ref{string}.  Such a
calculation is fully non-perturbative since the Feynman Path Integral
includes all possible interactions in the matrix element that we are
evaluating. In contrast to the real world, however, the calculations
are done with a non-zero value of the lattice spacing and a
non-infinite volume. In principle we must take $a \rightarrow 0$ and
$V \rightarrow \infty$ by extrapolation. In practice it suffices to
demonstrate, with calculations at several values of $a$ and $V$, that
the $a$ and $V$ dependence of our results is small, and understood,
and include a systematic error for this in our result.

\subsection{Algorithms}

\index{algorithms, lattice QCD}The Feynman Path Integral (Equation~\ref{fpi}) for gluons only becomes
\begin{equation}
\langle 0|\bm{\mathcal{O}}|0\rangle 
= \Frac{\int [dU]\, \bm{\mathcal{O}} e^{-S}}{\int [dU] \,e^{-S}}.
\end{equation}
To evaluate this integral we can generate random sets of $U$ 
fields on the lattice and work out the result:
\begin{equation}
\langle 0|\bm{\mathcal{O}}|0\rangle 
= \Frac{\sum\nolimits_{\alpha}\bm{\mathcal{O}}_{\alpha}e^{-S_{\alpha}}}
{\sum\nolimits_{\alpha}e^{-S_{\alpha}}}.
\end{equation}
$\{U\}_{\alpha}$ is a set of $U$ matrices, one for each link of 
the lattice, and is called a {\it configuration}. $\bm{\mathcal{O}}_{\alpha}$ is the 
value of $\bm{\mathcal{O}}$ on that configuration\index{configuration} (e.g.
the trace of a closed loop 
of $U$s). 
A set of configurations is an {\it ensemble}\index{ensemble}. 

\index{importance sampling}This is a very inefficient way of
working. If $S_{\alpha}$ is large for a particular configuration it
contributes very little to the result.  Instead it is better to
generate the configurations with probability $e^{-S}$. This is called
`importance sampling' since we preferentially choose configurations
with a large contribution to the integral.  If we have a set of
configurations so distributed then
\begin{equation}
\langle 0|\bm{\mathcal{O}}|0\rangle 
= \langle\bm{\mathcal{O}}\rangle 
= \frac{1}{N} \sum_{\alpha=1}^N \bm{\mathcal{O}}_{\alpha},
\end{equation}
i.e. the result simply becomes the ensemble average of the 
value of the operator $\bm{\mathcal{O}}$ evaluated on each configuration. 
The calculation then has a statistical uncertainty 
associated with it, which varies with the ensemble size, 
$N$, as $1/\sqrt{N}$.  

Several algorithms exist to generate an ensemble of configurations
with distribution $e^{-S}$. The Metropolis algorithm \index{Metropolis
algorithm}is the earliest and simplest, but shares several features
with later more sophisticated algorithms. The first step is to
generate a starting configuration, $\{U\}_1$, e.g. by setting all the
$U$ matrices to the unit $3\times 3$ matrix or by generating random
SU(3) matrices. The algorithm then sweeps round the configuration, one
$U$ matrix at a time. For each $U$ matrix a small change is proposed,
i.e. a random matrix close to the unit matrix is generated which could
multiply $U$.  The change in $S$ is calculated if this change to $U$
were to happen.  If $S$ is reduced, the change is accepted; if not, it
is accepted with probability $e^{-\Delta{S}}$ (by comparing
$e^{-\Delta{S}}$ to a random number between 0 and 1).  Once this has
been done for every $U \in \{U\}_1$ we have a new configuration,
$\{U\}_2$. We then repeat to obtain $\{U\}_3$ etc. Once we have an
ensemble we can do any number of different calculations (often called
`measurements') on it for different operators $\bm{\mathcal{O}}$. Ensembles are the
equivalent of experimental data sets created by collaborations of
theorists.  They are often stored for years and re-used many
times. Some ensembles are publicly available - see
http://qcd.nersc.gov/ and http://www.ph.ed.ac.uk/ukqcd/.

\begin{figure}[!h]
\centerline{\includegraphics[height=4.0cm]{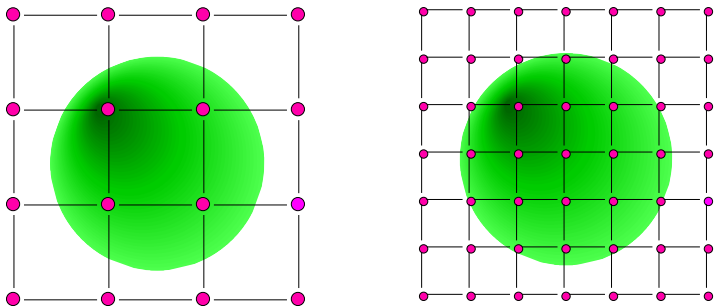}}
\caption{A given physical distance requires more lattice points to cover it
as $a$ is reduced.}
\label{critslow}
\end{figure}

An important point to note is that each member of an 
ensemble is generated from a previous member. The 
ensemble therefore has a (computer) time history. We have to worry 
about the `equilibration time' and the `decorrelation' 
(autocorrelation) time of the ensemble. The equilibration time is the number 
of sweeps required to reach a configuration typical of 
the distribution $e^{-S}$ that we are trying to create, 
i.e a configuration which has `forgotten' the starting 
configuration. The autocorrelation time is the number of 
sweeps it takes to generate a sufficiently different 
configuration that results can be considered statistically 
independent. The autocorrelation time can be determined from 
the sequence of results for $\bm{\mathcal{O}}$ and 
will depend on $\bm{\mathcal{O}}$. In general 
if $\bm{\mathcal{O}}$ is an operator with large extent, e.g. a closed loop of $U$ fields 
over many lattice sites, it will have a longer autocorrelation 
time than if $\bm{\mathcal{O}}$ is a small loop. This is because the changes 
to a configuration spread out randomly from a point, one step per sweep. 
As we try to reach smaller 
values of $a$, closer to the continuous space-time of 
the real world, we expect a phenomenon called `critical slowing-down'.  
This is because a given physical distance, say the size of 
a hadron, takes up many more lattice sites as $a$ gets 
smaller. For an ensemble to decorrelate on this physical 
distance scale then  requires more sweeps. This makes the numerical cost
of reducing the lattice spacing at fixed physical volume far worse
than the na\"{\i}ve $a^4$ (see \Figure{\ref{critslow}}). 

\subsection{Quarks on the lattice}

\subsubsection{The fermion doubling problem}

\index{quarks in lattice QCD}The inclusion of quarks in the lattice
QCD action causes several difficulties related to their fermionic
nature and makes lattice QCD calculations very costly in computer
time.

\begin{figure}
\centerline{\includegraphics[height=7.0cm]{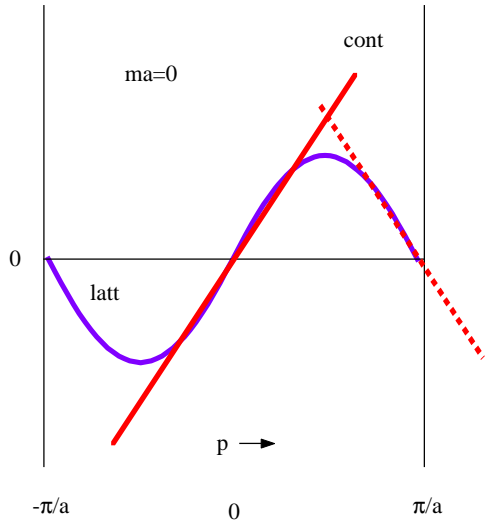}}
\caption[jgjg]{The doubling problem for lattice fermions. The sine
curve shows the lattice quark inverse propagator in 1-d. The straight
lines through $p$ = 0 (solid) and through $p = \pi/a$ (dotted) are
those for a continuum quark.}
\label{doubling}
\end{figure}

The so-called `fermion doubling' problem \index{fermion doubling
problem} is apparent even for free quarks, in the absence of any
interaction with the gluon field.  The continuum action for a single
flavor of free fermions is
\begin{equation}
S_f = \int d^4x \,\bar{\psi}(\gamma^{\mu}\partial_{\mu}+m)\psi.
\label{sfermcont}
\end{equation}
The obvious (so-called na\"{\i}ve) lattice discretisation gives
\begin{equation}
S_f^{\rm latt,naive} = a^4 \sum_x \left[\bar{\psi_x} \sum_{\mu=1}^4
\gamma_{\mu} \frac{\psi_{x+1_{\mu}} - \psi_{x-1_{\mu}}}{2a} + m
\bar{\psi_x}\psi_x \right].
\label{sqlatta}
\end{equation}
The problems become evident when we Fourier transform this and
compare the lattice inverse propagator:
\begin{equation}
G^{-1}_{\rm latt,naive}(p) = i \gamma_{\mu} \frac{\sin p_{\mu}a}{a} + m
\label{glattn}
\end{equation}
to that obtained in the continuum from Equation~\ref{sfermcont},
\begin{equation}
G^{-1}_{\rm cont}(p) = i \gamma_{\mu}p_{\mu} + m.
\label{gcont}
\end{equation}
The two are plotted for a massless quark in one-dimension in
\Figure{\ref{doubling}} over one lattice Brillouin zone (momenta beyond
$\pm \pi/a$ are equivalent to those in this range).  The lattice
result looks continuum-like around $p \approx 0$, where the inverse
propagator is close to zero.  The lattice inverse propagator is also
close to zero around $p \approx \pi/a$, however. Since $\pi/a$ and
$-\pi/a$ are periodically connected on the lattice, another
continuum-like line can be drawn at this point (with opposite slope to
the one at the origin). Thus in one-dimension, our lattice fermion
contains two continuum-like fermions rather than one! On a
4-dimensional lattice we have $2^4$ fermions instead of one. The 15
excess fermions are called doublers.  The doubling problem is clearly
a consequence of the fact that the {sine} function appears in
Equation~\ref{glattn} and this is because of the single derivatives in
the Dirac action for a relativistic fermion,
Equation~\ref{sfermcont}. For a scalar particle (Equation~\ref{sclr})
we would have a { cosine} instead, and no difficulty.

\subsubsection{Wilson quarks}

\index{Wilson quarks}There are several approaches to the doubling
problem. The most severe in terms of its effects, but currently the
most popular for a lot of applications, is the Wilson quark action. In
this the doublers are entirely removed, by adding a `Wilson' term to
the action which gives them a much larger mass than $ma$, so that they
drop out of the physics. The term added is a double derivative so
appears with an extra power of $a$ ($a^5$) in order to have the same
dimensions as the other terms in $S_f^{\rm latt}$
(Equation~\ref{sqlatta}):
\begin{eqnarray}
S_f^W &=& S_f^{\rm naive} - \frac{r}{2}a^5 \sum_x \bar{\psi}_x 
\mbox{\Large$\Box$} \psi_x \label{sw}, \\
\mbox{\Large $\Box$} \psi_x &=& \sum_{\mu=1}^4 \frac{{\psi}_{x+1_{\mu}} - 2 \psi_x + \psi_{x-1_{\mu}}}{a^2} ,
\end{eqnarray}
where $r$ is the Wilson parameter (almost always set to 1). The extra power
of $a$ in Equation~\ref{sw} means that the correspondence between the
lattice and continuum actions as $a \rightarrow 0$ is not changed.
However, if we look at the inverse propagator again, there is a
difference.
\begin{equation}
G^{-1}_W(p) = G^{-1}_{\rm naive} + \frac{2r}{a^2}\sum_{\mu=1}^4 \sin^2(p_{\mu}a/2).
\end{equation}
If we substitute for $G^{-1}_{\rm naive}$ from Equation~\ref{glattn} and 
expand out the $\sin$ function around $p \approx 0$ we get
\begin{equation}
G^{-1}_W(p) = i \gamma_{\mu}p_{\mu} + m + \frac{ra}{2}\sum_{\nu=1}^4 p_{\nu}^2.
\end{equation} 
Comparing this to the continuum form (Equation~\ref{gcont}) as $a
\rightarrow 0$, the $r$ term will disappear and a fermion of mass $m$
will have the right form.  If instead we look at the doublers, we must
expand around $p \approx \pi/a$.  If we call $\widetilde{p}$ the momentum
difference between $p$ and $\pi/a$ and consider the case where $p$ has
only one component close to $\pi/a$, and the others are close to zero,
then
\begin{equation}
G^{-1}(\widetilde{p}) 
= i \gamma_{\mu}\widetilde{p}_{\mu} + m + \frac{2r}{a} + \cdots\; .
\end{equation}
Now as $a \rightarrow 0$, the mass of the doubler, $m + 2r/a
\rightarrow \infty$.  The doublers at other corners of the Brillouin
zone pick up masses of $4r/a,\; 6r/a,\; 8r/a$ : check this as an exercise.
Thus we are assured that our quark action describes only the one quark
that we intended, but there is a price for this, as we shall see
below.

The Wilson quark action is converted to dimensionless units by a
rescaling $a^{3/2}\psi \rightarrow \psi$, leaving the quark mass
parameter as a mass in lattice units, $ma$ (previously called $m'$).
\begin{equation}
S_f^W = \sum_x \left\{\bar{\psi}_x \sum_{\mu} 
\left[(\gamma_{\mu}-r)\psi_{x+1_{\mu}} 
- (\gamma_{\mu}+r)\psi_{x-1_{\mu}}\right] 
 + (ma+4r)\bar{\psi}_x \psi_x \right\}.
\end{equation}
It is conventional to define a `hopping parameter' called $\kappa$
which is $1/(2ma+8r)$ and so $1/\kappa$ plays the r\^ole of the quark
mass.  $\psi$ is conventionally rescaled by $\sqrt{2\kappa}$ so that
$\kappa$ moves to multiply the terms connecting the $\psi$ field on
different sites (thus allowing `hops'). If we now couple in a gluon
field, the $\psi$ field will become a 3(color)${\times}$4(spin)
dimensional vector on each site. The gluon field must be included in
such a way as to keep the action gauge-invariant. From our earlier
discussion it is then obvious that $U$ matrices must be inserted as a
link between the $\psi$ and $\overline{\psi}$ fields when they are on
neighbouring sites. The Wilson quark action is then conventionally
written:
\begin{equation}
S_f^W = \sum_x 
\left\{ \kappa \left[\sum_{\mu} \bar{\psi}_x (\gamma_{\mu}-r) 
U_{\mu}(x)\psi_{x+1_{\mu}} 
- \bar{\psi}_{x+1_{\mu}}(\gamma_{\mu}+r)U_{\mu}^{\dag}(x)\psi_{x}\right] 
 + \bar{\psi}_x \psi_x \right\}.
\label{swilson}
\end{equation} 

The price we pay for using the Wilson quark action is that we break
explicitly the chiral symmetry of continuum QCD. This is a symmetry of
the derivative terms in $S_f$ (Equation~\ref{sfermcont}) which allows
us to rotate separately right- and left-handed components of the quark
field. The spontaneous breaking of this symmetry gives us a massless
pseudoscalar meson called the pion as a Goldstone boson and has other
important consequences for particle physics.  Chiral symmetry is
broken explicitly by a quark mass (so that the real pion is not
actually massless) but also, more seriously for the lattice, by the
Wilson term.  As $a \rightarrow 0$, chiral symmetry will be recovered,
but for real lattice calculations at non-zero~$a$, the lack of chiral
symmetry can cause difficulties for some calculations.

One surprising feature of Wilson quarks is that it is still possible
to get a massless pion even at non-zero $a$, when chiral symmetry is
broken. However, we have to search for the value of $1/\kappa$ at
which it occurs---it is not simply the point $1/\kappa = 8r$, as it
would be in the free theory, above.  Lattice calculations of the mass
of the pseudoscalar meson ($M_{\rm PS}$) must be done at various input
values of $\kappa$ (see Section~3) for a given ensemble.  A plot of
$M_{\rm PS}^2$ against $1/\kappa$ is then extrapolated to the point where
$M_{\rm PS}$ is zero.  The value of $\kappa$ at this point is called
$\kappa_{\rm critical}$ and is the point at which the bare quark mass in
the interacting theory is zero (but matrix elements will not
necessarily show chirally symmetric behaviour).  The bare quark mass
in lattice units, $ma$, at other values of $\kappa$ can then be taken
to be $(1/2\kappa - 1/2\kappa_{\rm critical})$.

Another problem for the Wilson quark action is the presence of large
discretisation errors. The na\"{\i}ve quark action has discretisation
errors proportional (at lowest power) to $a^2$ because (see
Equations~\ref{glattn} and \ref{gcont}) $\sin(pa)/a = p(1 - p^2a^2/6 +
\cdots )$. In the measurement of a hadron mass, the terms proportional
to $p^2a^2$ in the action will induce an error proportional to
$\Lambda^2a^2$ where $\Lambda$ is some typical momentum scale inside
the hadron in question, say 300MeV. For lattice spacing values we can
reach, around 0.1fm (= $(2{\rm GeV})^{-1}$ when $\hbar = c = 1$), this
gives an expected error of order 2\%. The Wilson term
(Equation~\ref{sw}) that we added, however, is proportional to $a$, so
that $S_f^W = S_f^{\rm cont} + {{O}}(a)$. Now hadron masses will have
an error of typical size $\Lambda a$, which could be 15\% at $a$ =
0.1fm.  One can extrapolate this error away by doing calculations at
several values of $a$ but the size of the extrapolation adds
uncertainty.

Instead, we can `improve' the quark action, by adding additional terms
to counteract the errors at any order in $a$. This is equivalent to a
higher order discretisation scheme for differential equations.  For
the Wilson quark action we can add the so-called clover term, making
the clover, or Sheikholeslami-Wohlerti, action:
\begin{equation}
S_f^{\rm clover} = S_f^W -\frac{iac_{\rm sw}\kappa r}{4} \sum_x \bar{\psi}_x \sigma_{\mu\nu}F_{\mu\nu}\psi_x.
\label{sclover}
\end{equation}
The standard discretisation of $aF_{\mu\nu}$ is as a set of 4
plaquettes arranged in a clover-leaf shape.  If the clover
coefficient, $c_{\rm sw}$, is chosen correctly then the clover action has
leading order errors proportional to $a^2$ again\index{clover quarks}.
It is in the correct choice of this coefficient that the difficulties
of discretising a field theory, as opposed to a standard differential
equation, appear. We are trying to match QCD with an ultraviolet
momentum cut-off of $\pi/a$ to QCD with an infinite momentum cut-off.
Gluonic interactions with gluon momenta between $\pi/a$ and $\infty$
in the continuum must be accounted for on the lattice by a
renormalisation of coefficients in the action. Thus the na\"{\i}ve
(tree-level) value of 1 for $c_{\rm sw}$ is renormalised by an amount
which depends on the QCD coupling constant at some momentum scale
around $\pi/a$. This momentum scale is typically quite large (for $a$
= 0.1fm it is 6GeV) so that a perturbative calculation of $c_{\rm sw}$
can work well.  $c_{\rm sw} = 1 + c_1 \alpha_s(\pi/a) + c_2
\alpha_s^2(\pi/a) + \cdots\,$.  In fact it has been shown that a lot of
the perturbative correction can be absorbed into a renormalisation of
the $U$ field by a factor called $u_0$, and this is called
tadpole-improvement (Lepage, 1993). Alternatively $c_{\rm sw}$ can be
determined within the lattice calculation itself (i.e.
non-perturbatively) by insisting that some continuum relationship,
broken by the discretisation errors, works on the lattice (Sommer,
1998).  For $c_{\rm sw}$ we can impose Ward identities from chiral
symmetry, for example.  This improvement programme for the lattice
action can be carried further at the cost of introducing more
coefficients that have to be determined by a match to continuum QCD.
However, this must be compared to the cost of {\it not} improving the
action, which requires calculations on very fine lattices to achieve
small enough discretisation errors for the accuracy we require and is
generally prohibitive.

\subsubsection{Staggered quarks}

\index{staggered quarks}Here we return to the na\"{\i}ve quark action
and ask, what was so bad about having 16 quarks instead of 1? If we
had 16 flavors of quarks of the same mass in Nature, the na\"{\i}ve
action might be fine. In fact we only have two quarks that might be
considered degenerate, $u$ and $d$. They both have masses of a few
MeV. Although we do not believe that their masses are the same, the
difference is much smaller than any other mass, and they are treated
as degenerate in most lattice calculations at present.

We can `thin' the degrees of freedom of the na\"{\i}ve lattice quark
action by removing the 4 spin degrees of freedom (which can be shown
to be multiple copies of the same thing).  The quark field, $\mychi$,
then becomes a 3(colors)$\times$1(spin) component object on a site and
the staggered (Kogut-Susskind) fermion action is:
\begin{equation}
S_f^S = \sum_x 
\bar{\mychi}_x \left\{ \frac{1}{2} \sum_{\mu} 
\eta_{x,\mu}\left(U_{\mu}(x)\mychi_{x+1_{\mu}} 
- U_{\mu}^{\dag}(x-1_{\mu})\mychi_{x-1_{\mu}}\right) 
+ ma \mychi_x \right\}.
\label{stagg}
\end{equation}
$\eta_x$ is $\pm 1$ according to the formula $\eta_{x,\mu} =
(-1)^k$ where $k={\sum_{\nu < \mu}x_{\nu}}$.  This action describes 16/4 = 4
quarks, now much closer to the real world, if we want to interpret the
doublers as flavors. We might hope that if the 4 flavors do behave as
4 copies of the same thing we can reduce their effect by a factor of
two or four (depending on how many degenerate flavors we want to
simulate) by multiplication with the required factor at appropriate
points (as we could in QCD perturbation theory).  The 4 spin degrees
of freedom for the 4 flavors are made from the 16 components of the
$\mychi$ field on a $2^4$ hypercube, which is a complication if we need
to separate out the flavors. The staggered action, however, has a
remnant of chiral symmetry which ensures the very desirable feature
that the quark mass (and the associated Goldstone boson pion mass)
vanish at $ma$ = 0.  This behaviour gives the added benefit of making
staggered quarks rather better behaved and computationally much faster
to work with than Wilson-type quarks.

\begin{figure}
\centerline{\includegraphics[height=4.0cm]{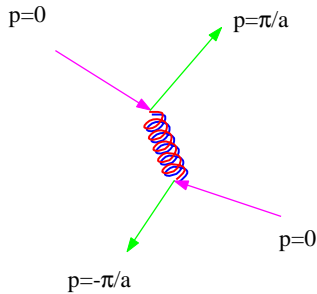}}
\caption{A flavor-changing interaction for staggered quarks on the lattice.}
\label{fcc}
\end{figure}

The down-side of staggered quarks is again the discretisation
errors. These are formally ${O}$$(a^2)$, just as for na\"{\i}ve
quarks, but some of the errors induce flavor-changing interactions and
so are rather dangerous. In practice they produce a larger than
expected effect for simple $a^2$ errors.  A quark with momentum around
0 can be scattered to one with momentum around $\pi/a$ i.e a doubler,
and therefore a different flavor, by the interaction of
\Figure{\ref{fcc}}.  One of the results of this is that the 16 different
pions (for 4 flavors) no longer have the same mass and only one of
them has a mass which vanishes as $ma \rightarrow 0$.  Improvement
terms have recently been developed which can be added to the action to
reduce these interactions to a much lower level, and the masses of the
different pions are then much closer together (Bernard, 2001, MILC
collaboration).  This makes the prospects for working with staggered
quarks in lattice QCD calculations much better, and a lot more work
with these quarks will certainly be done.

\subsubsection{Ginsparg-Wilson quarks}

\index{Ginsparg-Wilson quarks}A recent development has been a set of
quark actions which maintain chiral symmetry of the action while still
describing only one quark flavor, but at the cost of a very
complicated lattice discretisation of the continuum derivative. This
is then costly to implement. For example, in the domain-wall
formulation an additional 5th dimension is required whose length, in
principle, must go to infinity.  A lot of work is being done to
develop algorithms for these quark actions which may make them
feasible in the long-term.  In the meanwhile, they are already being
used for calculations that really need chiral symmetry at finite
lattice spacing, such as that of the CP-violating parameter in the $K$
system, $\epsilon'$.

\subsection{Algorithms for quarks}

Another problem with handling quarks in lattice QCD is that 
they are fermions, obeying the Pauli Exclusion Principle,
and therefore cannot be represented by ordinary numbers 
in a computer. We must do the quark functional integral 
by hand:
\begin{equation}
\int [dU]\, [d\psi]\, [d\bar{\psi}] \,e^{- S_g + \bar{\psi}M \psi} 
= \int [dU] \,{\det} M e^{- S_g}
\label{incq}
\end{equation}
where the form for the matrix $M$ depends on the quark formulation and
can be derived from the forms given above for the quark action
(Equations~\ref{swilson}, \ref{sclover} and \ref{stagg}). The QCD
action then becomes
\begin{equation}
S = \beta \sum_p 
\left(1 - \frac{1}{N_c}  \Re \Tr\left( U_p\right)\right) 
- \ln \left(\det M\right).
\end{equation}
We now generate ensembles of gluon fields (only) with importance
sampling based on this action. The standard algorithm for doing this
is called Hybrid Monte Carlo \index{Hybrid Monte Carlo}. The second
term is a very expensive one to include, because it requires frequent
calculations of $M^{-1}$ (various algorithms, such as Conjugate
Gradient exist to do this) and $M$ is a large matrix ($4 {\rm (for
Wilson)} \times 3 \times V \approx 2 \times 10^6$ on a side).  If this
term is missed out for expediency (so that the action is just $S_g$)
then we talk of using the `quenched approximation'. \index{quenched
approximation}Most calculations in the past have been quenched (and
most of the results I discuss later will be in the quenched
approximation) but recently calculations using the full QCD action
(`unquenched' or `with dynamical/sea quarks') have been attempted and
in the future we hope that the quenched approximation will become
redundant. We can think of the $\ln(\det M)$ term as giving rise to a sea
of quark/anti-quark pairs appearing and disappearing in the vacuum.
For every quark flavor for which we have a separate matrix $M$ we
should in principle include a term of the form $\ln(\det M)$ in the
dynamical quark action. However, it is only the production of light
($u,d,s$) quark/anti-quark pairs that we envisage having a significant
effect for most of the quantities that we calculate.  Dynamical
lattice calculations are then done with $N_f$ = 2 for $u,d$ dynamical
quarks or 2+1 if $s$ is included\index{dynamical quarks}.

Quarks must also be integrated out of the operators, $\bm{\mathcal{O}}$. 
For $\bm{\mathcal{O}}=(\overline{\psi}\psi)_y (\overline{\psi}\psi)_x$, 
the form mentioned earlier, which creates a meson 
at the point $x$ and destroys it at the point $y$, 
 then 
\begin{equation}
\int [dU] [d\psi] [d\bar{\psi}] \bar{\psi}^{u,a}_y \psi^{d,a}_y
\bar{\psi}^{d,b}_x \psi^{u,b}_x e^{-S} = \int [dU]
(M^{-1,u}_{x,y}[U])^{ab} (M^{-1,d}_{y,x}[U])^{ba}{\rm det} M e^{-S_g}.
\label{qfpi}
\end{equation}
$M^{-1}$ is the quark propagator from $x$ to $y$ on a given gluon
configuration, obtained by solving $Mx=b$ where $b$ is a vector with a
1 at $x$ (and a certain color and spin index) and $0$ everywhere else.
We have been explicit here about the flavor indices, which we have
taken as $u$ and $d$, although lattice calculations usually then
assume that $u$ and $d$ are degenerate and therefore the two $M^{-1}$
factors are the same. However, if the hadron actually does contain two
quarks of the same flavor then `disconnected' pieces containing
$M^{-1}_{x,x}$ will appear, as well as the `connected' pieces above.
The color indices, $a$ and $b$, are also explicit (and summed over)
and make $\bm{\mathcal{O}}$ gauge-invariant. The sums over spin indices have not been
made explicit because in this case they follow the color indices (but
see Section 2.6).  On an importance-sampled ensemble (either
quenched or unquenched) for this example we then have to calculate
${\rm Tr_{\rm color,spin}}(M^{-1,u}_{x,y})(M^{-1,d}_{y,x})$ on every
configuration and average over configurations.

Calculating $M^{-1}$ is computationally expensive and gets harder as
$M$ develops small eigenvalues, which happens as $ma \rightarrow 0$
(for staggered quarks) or $\kappa \rightarrow \kappa_{\rm crit}$ (for
Wilson or clover quarks). Thus, even in the quenched approximation, we
cannot actually calculate with quark masses close to those of real $u$
and $d$ quarks.  Instead we work with heavier quarks and perform
so-called chiral extrapolations to the chiral limit where $u$ and $d$
quarks would be (almost) massless.

\subsection{Relating lattice results to physics}

\begin{figure}
\centerline{\includegraphics[height=3.0cm,clip,trim=0 50 0 0]{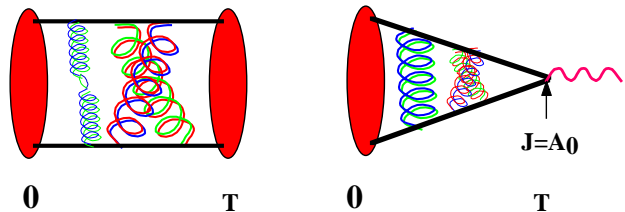}}
\caption{A graphical representation of two types of 2-point functions
calculated on the lattice.
Left, that for a hadron mass calculation; right, that for a decay constant.
}
\label{2pt}
\end{figure}

Above we have given an example for $\bm{\mathcal{O}}$, which includes the creation of
a valence quark and anti-quark at the point $x$ and their destruction
at the point $y$. This is a so-called hadron correlator or 2-point
function on the lattice \index{two-point function} since it simply has
a source and a sink, and is one of the simplest quantities to
calculate. It is shown pictorially at the left of \Figure{\ref{2pt}},
where the solid lines indicate the valence quark propagators, and the
blobs at the two ends indicate the creation and annihilation of the
meson. A baryon would of course have 3 valence quark propagator
lines. Usually we project onto specific values of $\mathbf{p}$ for the
hadron, so in Figure~\ref{2pt} we have suppressed spatial indices at
the source and sink and just refer to the time index, $0$ at the
source and $T$ at the sink. The figure shows how, as the valence
quarks propagate, they interact any number of times by exchange of
gluons. This is a pictorial representation of the fully
non-perturbative nature of a lattice QCD calculation. The interactions
include the production of dynamical quark/anti-quark pairs if a
dynamical calculation is being done.

The calculation of this 2-point function will enable the extraction of
the hadron mass (see below), for the hadron corresponding to the
$J^{PC}$ quantum numbers of $\bm{\mathcal{O}}$. We make different quantum numbers by
inserting $\gamma$ matrices between the $\psi$ and $\overline{\psi}$
fields in each piece of $\bm{\mathcal{O}}$. For example,
$(\overline{\psi}\gamma_5\psi)_x$ creates a pseudoscalar meson (such
as $\pi$) and $\overline{\psi}\gamma_i\psi$ a vector (such as
$\rho$). When the quark functional integral is done, as in
Equation~\ref{qfpi}, $\gamma$ matrices will appear between the two
$M^{-1}$ factors and appropriate sums over spin indices will have to
be done.

The blobs in Figure~\ref{2pt} indicate that we can use more
complicated forms for $\bm{\mathcal{O}}$ for a given hadron, e.g. the
$\psi$ and $\overline{\psi}$ fields do not both need to taken at the
point $x$. We can separate them spatially, either by inserting $U$
fields to keep $\bm{\mathcal{O}}$ gauge-invariant, or by fixing a
gauge to allow spatial separation without including $U$ fields. This
enables us to feed in information, or prejudice, about the relative
spatial distribution of the quarks in the hadron, i.e. its
`wavefunction'.\index{smeared operators} Each piece of
$\bm{\mathcal{O}}$ takes the form $\overline{\psi}_{x+r}\phi(r)\psi_x$
(suppressing the $U$ fields) where $\phi$ is some function of the
separation between $\psi$ and $\overline{\psi}$: it is known as the
`smearing' function and $\bm{\mathcal{O}}$ is then a smeared
operator. When the quark functional integral is done, factors of
$\phi$ will appear between the $M^{-1}$ factors. The factor of $\phi$
is absorbed at the source by solving $Mx=\phi$ for, say, the quark
(making a `smeared quark propagator') and $Mx=\delta$ for the
anti-quark (a `local quark propagator').  The two propagators are then
put together with an explicit insertion of $\phi$ at the sink.  Often
calculations measure separately hadron correlators with several
different smearing functions at both source and sink, enabling a more
precise determination of the hadron mass.

Another type of 2-point function is shown on the right of
Figure~\ref{2pt}.  In this case we create the hadron with a smeared
operator and destroy it with a local operator. This is a
`smeared-local' or `smeared-current' correlator, since the quantity
that we can extract from this is the matrix element of the appropriate
current operator, $J$, between the vacuum and the hadron.  For
example, this is used to calculate the decay constant, $f_{\pi}$,
related to the vacuum to $\pi$ matrix element of the axial vector
current (denoted by its time component, $A_0$, in Figure~\ref{2pt}).
This couples to the $W$ particle and mediates the purely leptonic
decay of a $\pi$ meson. See the Lagrangian for the weak interactions
in (Rosner, 2002), but note that the $W$ particle is not included
explicitly in lattice QCD calculations.  $\bm{\mathcal{O}}$ in this case then takes
the form $(\overline{\psi}_{x+r}\gamma_5\phi(r) \psi_x)
(\overline{\psi}\gamma_0\gamma_5\psi)_y$, where the first factor
creates the pion with a smeared operator at $x$ and the second
destroys it with the time component of the local axial vector
current. The quark functional integral converts this to the same type
of quantity, with two factors of $M^{-1}$, that we discussed above.

\begin{figure}[!h]
\centerline{\includegraphics[height=4.0cm,clip,trim=0 60 0 0]{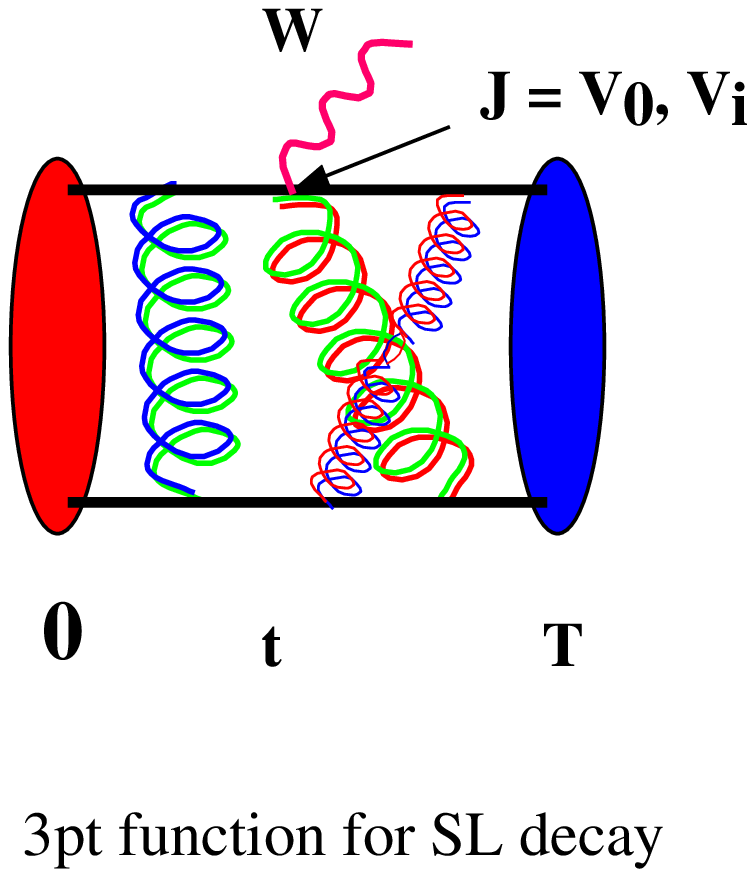}}
\caption{A graphical representation of a 3-point function (for semileptonic
decay) calculated
on the lattice.}
\label{3pt}
\end{figure}

\Figure{\ref{3pt}} shows a lattice 3-point function \index{three-point
function}appropriate to the semi-leptonic decay of a hadron. One of
the valence quark lines emits a $W$ and changes to a different
flavor. A new hadron is then formed with the spectator quark. The
emission of the $W$ can be represented by the insertion of a current
on one of the valence quark lines. The Figure shows a vector current
(with temporal component $V_0$ and spatial component $V_i$) which
contributes to the decay of a pseudoscalar meson to a pseudoscalar
meson (e.g. $B \rightarrow D$).  We then have a (smeared) source and
sink at $0$ and $T$, and a (local) current insertion at $t$, i.e. 3
points. When the quark functional integral is done there will be 3
factors of $M^{-1}$, one for the original valence quark which decays
(from 0 to $t$), one for the final valence quark (from $t$ to $T$) and
one for the spectator (from 0 to $T$).  In fact the most efficient way
to do this calculation is to solve for the final valence quark
propagator from $T$ to $t$, taking as a source the spectator quark
propagator from 0 to $T$.

\section{Lattice QCD calculations}

\subsection{The steps of a typical lattice calculation}

\noindent{\bf Step 1}

\noindent
A volume and a rough lattice spacing are chosen. A volume of $(3{\rm
fm})^3$ is considered to be large enough not to `squeeze', and
therefore distort, typical hadrons placed on it. The time extent is
usually taken as twice the spatial size since masses etc are extracted
from the time dependence of hadron correlators (see below).  The
selection of the lattice spacing is a trade-off between getting close
to the continuum limit (and therefore small discretisation errors) and
the cost of the calculation, which grows as some large power of
$a^{-1}$. Improvement of the action, discussed above, helps here by
giving small discretisation errors on coarser lattices. Lattice
spacings around 0.1fm are reasonable on both counts. From experience
we know roughly what value of the bare QCD coupling constant to take
in the gluon part of the QCD action to achieve various values of $a$
(determined {\it after} the calculation, see below).  However, the
quark contribution to the action affects this also, and we have much
less experience with this.  A $(3{\rm fm})^3 \times 6{\rm fm}$
lattice with $a \approx$ 0.1fm requires $(30)^3 \times 60$ sites.

\noindent{\bf Step 2}

\noindent
A quark formulation, number of quark flavors, and masses in lattice
units, $ma$, are chosen for the quark part of the QCD action. Again we
have a trade-off between trying to take realistically small masses for
the $u$ and $d$ quarks, and the cost. Again we do not know what the
quark mass actually is until after the calculation, when we have
calculated the masses of hadrons containing that quark.  Recent
calculations have been able to take dynamical quark masses down to the
$s$ quark mass and some have gone further; future calculations need to
reach much smaller masses than this. Extrapolations to $u$ and $d$
quark masses will continue to be necessary, however (see step 8).
Some interpolation will always be necessary too since the masses
chosen will inevitably not be exactly correct, e.g. for the physical
strange quark mass.

\noindent{\bf Step 3}

\noindent
An ensemble of gluon configurations must then be generated using
importance sampling with $e^{-S}$. As discussed above, dynamical
quarks appear implicitly through the quark determinant.

\noindent{\bf Step 4}

\noindent
Quark propagators are calculated on each gluon configuration of the
ensemble by inverting the quark matrix, $M$, to make the `valence'
quarks inside the hadron.  Where they are supposed to have the same
flavor as the dynamical quarks, they should have the same mass in
lattice units, $ma$.  However, we can also calculate valence quark
propagators for quarks with different mass from the dynamical quarks,
and perform separate extrapolations in valence and dynamical quark
masses. This is sometimes useful and particularly so if there is a
very limited set of dynamical quark masses. It is known as the
partially quenched approximation (PQA).

\noindent{\bf Step 5}

\noindent
The quark propagators are then put together in various combinations to
form hadron correlators (see the discussion of the form taken for
operators, $\bm{\mathcal{O}}$, above) which are then averaged over all the
configurations in the ensemble.  We are concentrating here on
operators $\bm{\mathcal{O}}$ which are related to quark-based hadrons but gluonic
operators can also be measured on the ensemble and averaged in the
same way.

\noindent{\bf Step 6}

\begin{figure}
\centerline{\includegraphics[height=7.5cm]{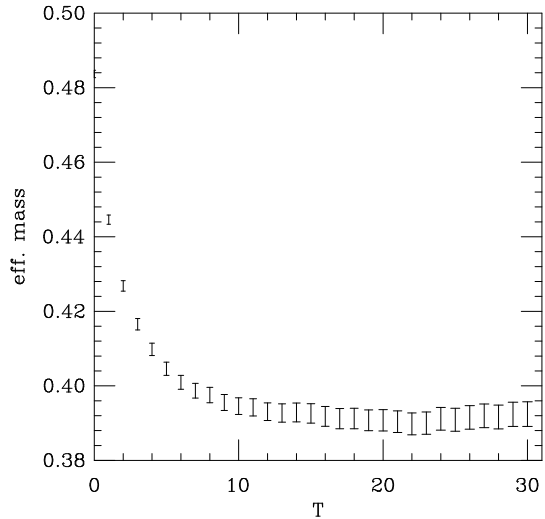}}
\caption{The effective mass of a $1^{--}$ $b\overline{b}$ ($\Upsilon$)
correlator, calculated from a lattice 2-point function with local
source and sink.}
\label{upseffmass}
\end{figure}

\noindent
The hadron correlators are fitted to their expected theoretical form
to extract hadron masses and matrix elements. For the 2-point function
described for the spectrum, the ensemble average of the product of
smeared quark propagators described above gives us the vacuum
expectation value of a hadron correlator, $\langle 0 | H^{\dag}(T)
H(0) | 0 \rangle$ (see Equation~\ref{fpi}). The hadron
creation(destruction) operator can create(destroy) from the vacuum all
the hadron states which have the same $J^{PC}$ quantum numbers as the
operator.  For example, if the operator has the quantum numbers of a
pseudoscalar meson containing $u$ and $d$ quarks, the $\pi$ and all
its radial excitations can be created(destroyed).  The amplitude, $A$,
with which a particular state is created or destroyed depends on the
overlap with that state of the operator used, i.e. the smearing
function, at source or sink. Thus we obtain
\begin{equation}
\langle 0 | H^{\dag}(T) H(0) | 0 \rangle = \sum_n \frac{A_{{\rm src}, n}
A_{{\rm snk}, n}}{2E_n} e^{-E_nT}
\label{expform}
\end{equation}  
where the factor $e^{-E_nT}$ arises because the two hadron operators
are offset by a time distance $T$ in Euclidean space, and $E_n$ is the
energy of the $n$th state. The states which dominate the fit,
especially at large values of $T$, are those with lowest energy; if a
projection on zero momentum has been done, these will be the states
with lowest mass.  Often we are interested in the one state with
lowest mass, the ground state (the $\pi$ in the example above), and
then try to design a good smearing function to have large overlap with
that state, and very small overlap with its radial excitations.  In
that case fits can be done in which data at small values of $T$ are
thrown away and only a single exponential is used in the fit. The
extent to which this works can be gauged by plotting the `effective
mass', the log of the correlator at time $t$ divided by $t$. If one
state completely dominates the fit, a constant result is obtained as a
function of $t$ - the effective mass is said to `plateau'. The plateau
value is the ground state mass. \Figure{\ref{upseffmass}} shows the
result for the effective mass\index{effective mass} of the correlator
for an $\Upsilon$ particle (see Section 4) calculated on the
lattice. A clear plateau is seen but only for $t >$ 15.  For smaller
$t$ the correlator clearly contains excitations of higher mass,
because no smearing was used in this case.

The best calculations use several different smearing functions at
source and sink and perform simultaneous multi-exponential fits of the
type in Equation~\ref{expform}. If the masses of several states can be
obtained from the fit the reliability of the ground state mass is
increased.  It should also be pointed out that correlated fitting
techniques must be used since the correlators at adjacent times are
not statistically independent of each other.

For the 2-point function used to calculate decay constants, the
amplitude with which the hadron is destroyed at the sink is the vacuum
to hadron matrix element of the current.
\begin{equation}
\langle 0 | J(T) H(0) | 0 \rangle = \sum_n \frac{A_{{\rm src}, n} \langle 0
| J | n \rangle}{2E_n} e^{-E_nT}
\label{decayc}
\end{equation}
and $\langle 0 | A_0 | \pi (\mathbf{p}=0) \rangle = f_{\pi}m_{\pi}$. To
isolate the part proportional to the decay constant requires dividing
the total amplitude of the ground state exponential by
$A_{{\rm src},n=g.s.}$. This can be obtained from a fit of the type in
Equation~\ref{expform}, if the same smearing function is used at
source and sink so that $A_{{\rm src},n} = A_{{\rm snk},n}$.

For the 3-point function we have two sets of hadrons with different
flavor quarks, separated by a current insertion.
\begin{equation}
\langle 0 | H'^{\dag}(T) J(t) H(0) | 0 \rangle = \sum_n \sum_m
\frac{A_{{\rm src}, n} A_{{\rm snk},m} \langle m | J | n \rangle}{2E_n2E_m}
e^{-E_nt}e^{-E_m(T-t)}
\label{3ptfit}
\end{equation}
where $n$ runs over hadrons with the quantum numbers of the operator
at 0 and $m$, those of the operator at $T$.  Again the matrix element
of interest, that of the current between two hadrons (usually for the
ground states in the two cases), can be obtained by dividing out the
amplitudes at source and sink from two separate 2-point fits for the
two different hadrons of the kind in Equation~\ref{expform}.

\noindent{\bf Step 7}

\noindent
It is now possible to determine what the lattice spacing was in the
simulation. This then sets the single dimensionful scale so that
everything can be converted to physical units (GeV) from lattice
units.  The lattice spacing is determined by requiring one
dimensionful quantity to take its real world value. Usually a hadron
mass is chosen, because these are easiest to determine on the lattice,
but it should not be one whose mass depends strongly on valence quark
masses to be determined in the next step (see below) otherwise a
complicated iterative tuning procedure will result. The most popular
quantity to use at present is known as $r_0$, a parameter associated
with the potential between two infinitely heavy quarks. It is
extracted from the energy exponent of a gluonic operator (the closed
loop of Figure~\ref{string}), so can be precisely determined and does
not contain any valence quark masses.  The only problem is that it is
not an experimentally accessible quantity, and we rely on potential
model results to give a phenomenological value, estimated to be
0.5fm. Another quantity frequently used is the mass of the $\rho$
meson, obtained by chiral extrapolation to the point where the $\pi$
meson mass, and therefore the $u$,$d$ quark mass, is (almost)
zero. The chiral extrapolation, however, can produce large errors. A
better quantity is the orbital excitation energy, i.e the splitting
between P states and S states, in $b\overline{b}$ or $c\overline{c}$
systems, since these don't contain light quarks and this splitting is
even insensitive to the heavy quark mass. (The treatment of heavy
quarks on the lattice will be discussed in Section 4.)

\noindent{\bf Step 8}

\noindent
The step above yields all hadron masses in GeV. However, before we can
compare to experiment we must tune the quark masses.  This requires
calculations at several different values of the bare quark masses in
an appropriate region. For each quark mass we then select a hadron
whose mass will be used for tuning (and is therefore not
predicted). For that hadron we interpolate/extrapolate the results to
find the bare quark mass at which that hadron mass is correct.  The
masses of other hadrons containing that quark are then predicted if we
interpolate/extrapolate those masses to the same quark mass, or
combination of quark masses. In the process we learn about the
dependence of hadron masses on the quark mass and this can be useful
theoretical information. The hadrons used for tuning should be
low-lying states with accurate experimental masses which can be
calculated precisely on the lattice. The $\pi$ mass is usually used to
fix the $u$, $d$ mass (taken to be the same), although sometimes the
approximation $m_{\pi}=0$ is used. The mass of the $K$, $K^{{\myast}}$, or
$\phi$ can be used to fix the $s$ quark mass. The $K$ or $K^{{\myast}}$
obviously require the $u$ and $d$ masses to have been fixed. The
dimensionless ratio of the $K^{\myast}$ to the $K$ mass can also be used, and
this is then less dependent on the quantity used to fix the lattice
spacing.  For the $c(b)$ quarks, the $D(B)$, $D_s(B_s)$ or $\psi
(\Upsilon)$ systems are convenient ones.

The interpolation/extrapolation of hadron masses as a function of bare
quark masses is a relatively simple procedure in the quenched
approximation. Then there is no feedback from the quark sector into
the gluon sector. We can create gluon field configurations at a fixed
value of the lattice spacing (as determined, for example, from a
purely gluonic quantity such as $r_0$) and measure hadron masses at
many different quark masses on those configurations. The issues are
then the correlations between results at different quark masses that
must be taken into account and the spurious non-analytic behaviour in
quark mass that can arise in the quenched approximation in
extrapolations to $u$ and $d$ masses (`quenched chiral logarithms').

When we include dynamical quarks in the calculation, the effects of
the quark determinant at a particular quark mass feed into the gluon
field configurations. Results at different dynamical quark masses then
represent a completely new calculation, generating a new ensemble of
gluon configurations with statistically independent results. The
interpolations/extrapolations in quark mass take on a new dimension
and there are subtleties associated with how to do this. Some groups
have chosen to generate configurations at fixed bare coupling constant
and various dynamical bare quark masses. Then the lattice spacing will
vary with quark mass and extrapolations in quark mass must be done in
lattice units, before fixing the lattice spacing at the end. I believe
a more satisfactory approach from a physical perspective is to adjust
the bare coupling constant at different bare quark masses so that the
lattice spacing remains approximately the same (as determined from
$r_0$, for example). This then allows interpolations/extrapolations
for physical hadron masses, and a better picture of the physical
dependence of quantities on the presence of dynamical quarks. Several
groups have also carried out this procedure.

In all of these approaches we must extrapolate to reach the physical
$u/d$ mass region, and so we need to know the appropriate functional
form for this extrapolation. This can be derived for light enough
$u/d$ mass using an effective theory of Goldstone pions called chiral
perturbation theory. This shows that logarithmic behaviour of
quantities as a function of the $\pi$ mass (the variable representing
the $u/d$ quark mass) should be present in general as well as simple
power-law behaviour.  These `chiral logarithms' will only show up at
rather small quark masses ($m_{u,d}\stackrel{<}{\sim} m_s/4$) and so
it is important for dynamical simulations to reach quark masses low
enough to be able to match on to this behaviour and extrapolate down.

\vfill
\noindent{\bf Step 9}

\noindent
The calculation needs to be repeated at several values of the lattice
spacing to check that the dependence of physical results on the
lattice spacing is at an acceptable level and/or to extrapolate to the
continuum limit $a = 0$. Extrapolations again obviously require
knowledge of an appropriate functional form.

\vfill
\noindent{\bf Step 10}

\noindent
Compare to experiment or give a prediction for experiment!

\newpage
\noindent{\bf Concluding remark }

\noindent
Above we have described an ideal situation. Lack of computer power has
meant compromising on one or more aspects in existing calculations.  A
lot of calculations have used the quenched approximation. More recent
dynamical calculations have used heavy dynamical masses on rather
coarse and sometimes rather small lattices.  These difficulties should
be overcome in the next few years and this will represent a huge
improvement in the reliability of lattice results.

\subsection{Control of lattice systematic errors}

We aim for errors of a few percent from future lattice calculations.
This requires both improved statistical errors in general and good
control of systematic errors.   Improved statistical accuracy is
obtained by generating larger ensembles of configurations with a cost
proportional to the square of the improvement.   Improved
systematic accuracy requires theoretical understanding of the sources
of error and how to remove them. It is this understanding, described
below, that has been responsible for the development of good lattice
techniques and the convergence of lattice results in the quenched
approximation through the late 1990s.  This must be carried further in
the next phase of dynamical simulations to reach the goal of providing
quantitative tests of QCD and input to experiment.

\vspace{-1ex}
\subsubsection{Discretisation errors}

\index{discretisation errors}As discussed earlier, these arise from
errors in the lattice form of the Lagrangian, and operators $\bm{\mathcal{O}}$,
compared to the continuum versions. Lattice results, even when
converted to physical units, have some dependence on $a$. This will be
as a power series in $a$, starting at $a^n$. As discussed earlier, $n$
= 1 if the Wilson quark action is used, 2 for the clover quark action
and 2 for the staggered quark action. $n$ is also 2 for the Wilson
plaquette gluon action of Equation~\ref{sglatt}.  We expect the size
of the $a$ dependence to be controlled by a typical momentum scale
relevant to the quantity being calculated. Quantities sensitive to
shorter distances than others will be more susceptible to
discretisation errors, even though the value of $n$ depends only on
the action used.  Improved gluonic and quark actions are available in
which higher order terms are added to $\cal{L}$ to increase $n$, and
therefore reduce the $a$ dependence, and these can be tested for their
efficacy in the quenched approximation.  The systematic improvement
method is known as Symanzik improvement (Gupta, 1998).

\begin{figure}
\centerline{\includegraphics[bb= 0 0 4096 4096,width=9cm,clip,trim=0 40 0 50]{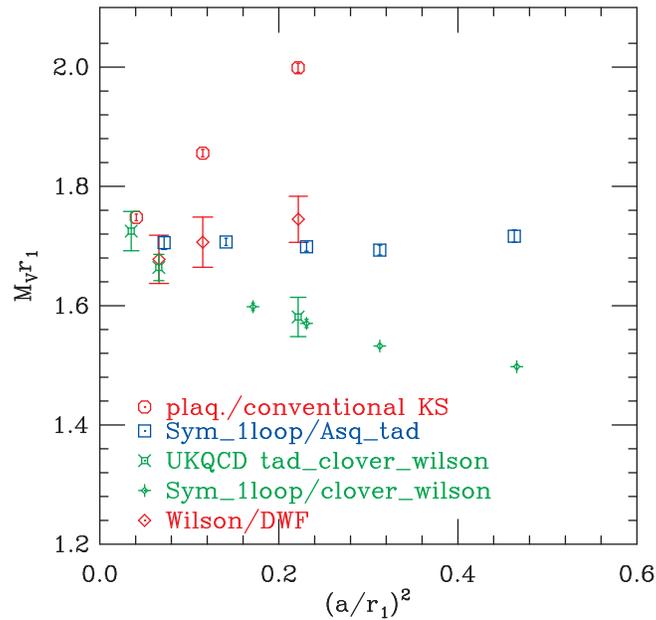}}
\caption[uioj]{A scaling plot in the quenched approximation for the
vector meson at a quark mass such that the pseudoscalar meson has mass
$m_{\pi}r_1 = 0.807$. The vector meson mass is given in units of
$r_1$, a variant of $r_0$, where $r_1 \approx$ 0.35fm $= 0.57 {\rm
GeV}^{-1}$.  It is plotted versus the square of the lattice spacing,
also given in units of $r_1$. The squares and fancy diamonds use an
improved gluon action; the others use the Wilson plaquette action. The
quark actions used are: circles, staggered (Kogut-Susskind); squares,
improved staggered; fancy squares and fancy diamonds, tadpole-improved
Wilson (clover); diamonds, Ginsparg-Wilson (domain wall). (Toussaint,
2002)}
\label{milc-rho}
\end{figure}

\Figure{\ref{milc-rho}} shows a scaling plot of the vector meson mass
(the $\rho$, except that the quark mass is heavier than the real $u,d$
mass) in GeV versus the lattice spacing for various quark actions
(Toussaint, 2002).  Some of the calculations use an improved gluon
action, with discretisation errors reduced beyond ${O}$$(a^2)$,
but others use the Wilson plaquette action. There is very little
difference between these (compare fancy diamonds and squares) so that
most of the difference arises from the quark action used. A variant of
$r_0$, called $r_1$, is used to set the lattice spacing so the vector
mass and scale are given in units of $r_1$. The plot shows results for
clover quarks (improved Wilson quarks), staggered quarks, improved
staggered quarks and Ginsparg-Wilson (domain wall) quarks. The last
two formulations, which are both improved to remove ${O}$$(a^2)$
errors show an impressively flat line, i.e. very little $a$ dependence
for this quantity. The clover quarks shown here have a clover
improvement coefficient (see Section 2.4.1) chosen using
tadpole-improvement. This reduces the $a$ dependence of Wilson quarks
to $\alpha_s a$ but it is clearly still visible. A non-perturbative
determination of the clover improvement coefficient can reduce the $a$
dependence further to ${O}$$(a^2)$, and then this formulation
looks rather better.  Notice the large discretisation errors visible
for unimproved staggered quarks, despite the fact that the errors are
${O}$$(a^2)$ (and results therefore lie on a straight line in the
Figure). Provided that all the different quark formulations have been
fixed to the same physical quark mass, all the results for the vector
meson mass should agree in the $a \rightarrow 0$ limit. This does seem
to be true, within the statistical errors shown.

\vspace{-1ex}

\subsubsection{Finite volume}

\index{finite volume errors}Lattice results will be distorted if the
space-time box in which the calculation is done is too small to
adequately represent the infinite space-time volume of the real
world. For large enough volumes the error should be exponential in the
lattice size, $\propto e^{-ML}$, for a lattice of size $L$ in physical
units. This means that it is possible to reduce finite volume errors
rapidly to zero by taking large enough volumes. The lightest particle
is the $\pi$, so this sets the volume required as we reduce the $u,d$
quark masses to their physical values. For $u,d$ quark masses of
$m_s/4$, $m_{\pi}L > 5$ for $L >$ 3fm, giving a finite volume error
of less than 1\%. Most recent lattice calculations have used volumes
of this size, although there has been little systematic dependence of
the volume dependence of results.

\subsubsection{Matching hadronic matrix elements to the continuum} 

\index{matching lattice matrix elements}The calculation of hadronic
matrix elements of various currents, $J$, on the lattice is discussed
for 2- and 3-point functions in Section~2.6.  An important point is
that these depend in general on how QCD has been regularised and a
finite renormalisation is then required to convert lattice results to
those appropriate to a continuum scheme (such as
$\overline{MS}$). Since lattice QCD and continuum QCD differ in the
ultra-violet (for momenta greater than $\pi/a$), this renormalisation can be
calculated in perturbation theory, by matching the matrix elements of
$J$ between quark states. We usually need several lattice currents to
make up the continuum current and a mixing and matching calculation
must be done.
\begin{eqnarray}
J_{\rm cont} &=& Z_0 J_{\rm latt}^{(0)} + a Z_1 J_{\rm latt}^{(1)} + \cdots \nonumber \\
Z_i &=& 1 + c^{(1)}_i \alpha_s(2/a) + c^{(2)}_i \alpha_s^2(2/a) +\cdots 
\end{eqnarray}

Lattice perturbation theory is done in the same way as continuum
perturbation theory, in terms of the field $A_{\mu}$ and including
gauge-fixing and ghost terms, if necessary. Relatively little lattice
perturbation theory has been done up to now and few results exist
beyond ${O}$$(\alpha_s)$.  This leaves errors of
${O}$$(\alpha_s^2)$, 5--10\% if we take a scale for $\alpha_s$ of
$2/a$ at $a{=}0.1$fm.  Higher order calculations will be required to
reduce this to the required level of 2--3\%, and techniques are being
developed to do this. It is also sometimes possible to fix the
normalisation of lattice currents non-perturbatively using symmetry
arguments or to match numerically between lattice and continuum
$MOM$-type schemes.  In whatever way it is done, the matching of
lattice matrix elements to the continuum is a lot of work and an area
where improvements are still necessary.

\subsubsection{Unquenching}

The neglect of dynamical quarks in the quenched approximation is
obviously wrong, but how wrong? For many years systematic errors from
the quenched approximation were obscured by the size of the
statistical and discretisation errors. Now improved quenched
calculations are showing internal inconsistencies and disagreement
with experiment which we believe will be removed once realistic
dynamical calculations can be done. \index{unquenching}

\begin{figure}
\centerline{\includegraphics[height=8.0cm]{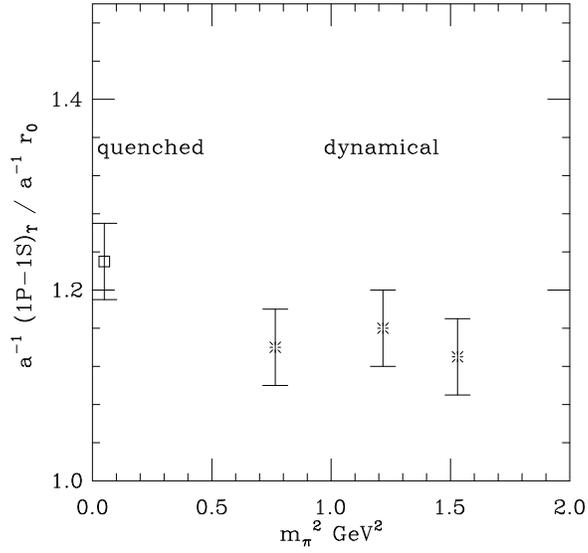}}
\caption{The ratio of inverse lattice spacings, $a^{-1}$, obtained
from the orbital excitation energy, the splitting between 1P and 1S
states, in the $\Upsilon$ system and from $r_0$. Results are given for
quenched simulations and for dynamical simulations using two flavors
of dynamical quarks at three different values of the quark mass, all
heavier than $m_s$, indicated by the square of the corresponding pion
mass along the $x$ axis. (Marcantonio, 2001, UKQCD collaboration)}
\label{aqunq}
\end{figure}

\begin{figure}
\centerline{\includegraphics[height=7.0cm]{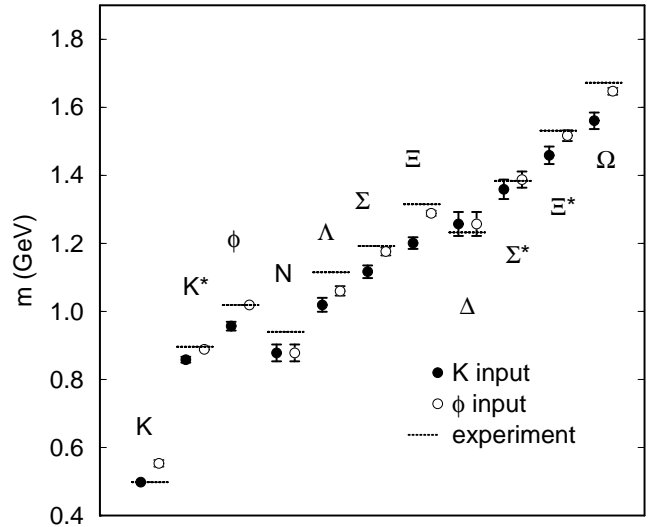}}
\caption{The spectrum of light mesons and baryons obtained in the
quenched approximation after extrapolation to $u,d$ quark masses and
to the continuum limit. The $\rho$ and $\pi$ masses are missing since
they were used to fix the lattice spacing and $u,d$ masses. Results
are compared using the $K$ or the $\phi$ to fix the strange quark mass
and disagreement between the two is seen. The size of the discrepancy
with experiment depends on this and varies between hadrons, but is at
the level of 10\%. (Aoki, 2000, CP-PACS collaboration)}
\label{cppacs}
\end{figure}

One effect expected in the quenched approximation is the incorrect
(too fast) running of the coupling constant from one scale to another
because of the absence of $ g \rightarrow \overline{q}q \rightarrow g$
pieces in the vacuum polarisation to give quark screening of the color
charge. From this we might expect that the determination of the
lattice spacing would depend on the quantity used to fix it, since
different quantities will be sensitive to different distance/momentum
scales and these will not be connected correctly by the running of
$\alpha_s$ in the quenched approximation.  (Using a quantity to fix
$a$ is equivalent to fixing the QCD coupling constant at the momentum
scale relevant to that quantity).  This is indeed found and
illustrated by the quenched point in \Figure{\ref{aqunq}}. Likewise
hadron masses depend on the hadron used to fix the quark mass. Then if
a set of hadron masses is studied, sensitive to a range of scales and
containing different combinations of quarks, errors will show up (see
\Figure{\ref{cppacs}} (Aoki, 2000, CP-PACS collaboration)).

The quenched approximation also does not allow the decay of particles
where this requires the production of a $\overline{q}q$ pair from the
vacuum, e.g $\rho \rightarrow \pi\pi$. Once dynamical quarks are light
enough for this to happen, it will in fact be difficult to determine
$m_{\rho}$ since we will obtain instead the lighter mass of the
two-pion system.  It is then important in dynamical simulations to use
hadrons which are stable in QCD, or have very narrow widths, to fix
the quark masses in the QCD action.

\begin{figure}
\centerline{\includegraphics[height=8.0cm]{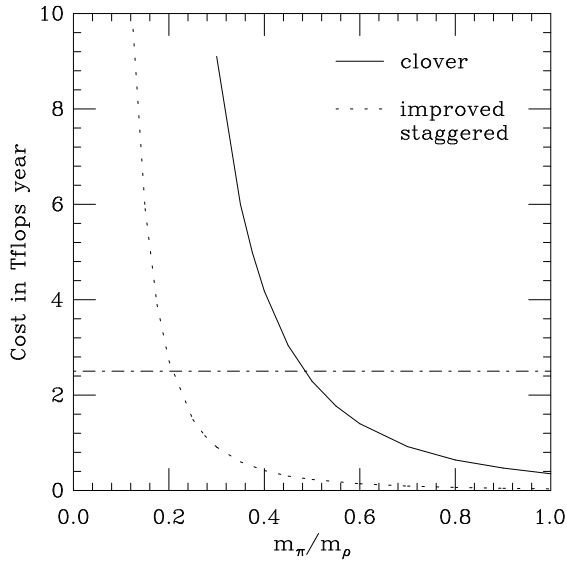}}
\caption{The computer cost in teraflop-years of generating 500 
$30^3\times 60$ configurations with $a =$ 0.1fm and a 
dynamical quark mass which gives the ratio of pseudoscalar to 
vector meson masses along the $x$ axis. Clover and improved 
staggered quarks are compared, assuming the same scaling behaviour
as $(m_{\pi}/m_{\rho})^3$. The straight line shows what is 
possible in 6 months on a 5Tflops computer.}
\label{costdyn}
\end{figure}

It has been stressed that the numerical cost of unquenched
calculations is very high. It increases very rapidly as $a$ is reduced
at fixed physical volume and as $m_{u,d}$ is reduced, although the
exact scaling behaviour is not completely clear. \Figure{\ref{costdyn}}
estimates the cost of generating an ensemble of 500 gluon
configurations on an $L^3\times T$ lattice with $L$ = 3fm and $T=2L$
at a lattice spacing, $a$ = 0.1fm, as a function of the $u,d$
dynamical quark mass. The $x$ axis is plotted as the ratio
$m_{\pi}/m_{\rho}$ where the $\pi$ and $\rho$ are the pseudoscalar and
vector mesons made with valence quarks of the same mass as the
dynamical quarks. The real world has $m_{\pi}/m_{\rho}$ = 0.2. For
$m_{u,d} = m_s$ the ratio is 0.7, for $m_{u,d} = m_s/2$, 0.55 and for
$m_s/4$, 0.4.  For $m_s/2$ the ratio is obtained from the $K$ and
$K^{\myast}$ masses.  For $m_s$ and $m_s/4$ some arguments must be made about
the scaling of hadron masses with quark masses because, for example,
no pure $s\overline{s}$ pseudoscalar meson exists.  The cost varies
here as $(m_{\pi}/m_{\rho})^3$, which is based on estimates from
simulations (LAT2001). Figure~\ref{costdyn} compares the cost for clover quarks
and improved staggered quarks, again based on simulations at one quark
mass, and using the same scaling formula. The cost advantage of
improved staggered quarks is clear on this plot. One disadvantage is
that the algorithm generally used for two flavors of dynamical
staggered quarks is not exact, unlike that for clover. This means that
there are systematic errors, rather like discretisation errors, which
increase with the computer time step, $\epsilon$, which is used to
generate one gluon configuration from the previous one. Checks must to
be done to make sure that this systematic error is at an acceptable
level and/or an extrapolation to $\epsilon$ = 0 must be done.

Recent unquenched calculations, albeit with rather heavy dynamical
quark masses, have shown encouraging signs that systematic errors from
the quenched approximation are being overcome.  Figure~\ref{aqunq}
shows that the ratio of $a^{-1}$ values obtained from two different
quantities is closer to 1 on dynamical configurations (using two
flavors of dynamical quarks with a mass around $m_s$) than it was on
quenched configurations (Marcantonio, 2001, UKQCD collaboration).
From this we can hope that with 2 dynamical light quarks and a
dynamical strange quark there will be only one value of the lattice
spacing, corresponding to the one dimensionful scale of QCD in the
continuum.

\begin{figure}
\centerline{\includegraphics[height=7.0cm]{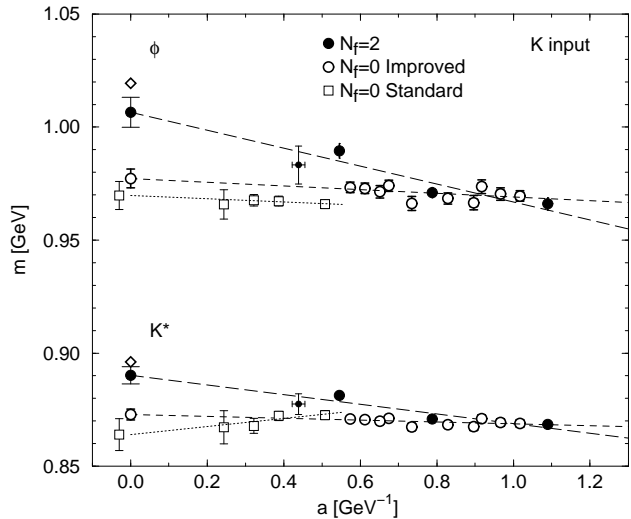}}
\caption{ The masses of the $\phi$ and $K^{\myast}$ mesons as a function of
lattice spacing, $a$, for quenched simulations and those using 
two flavors of dynamical quarks. The $K$ meson is used to fix the 
strange quark mass. The experimental results are indicated by 
diamonds at $a =$ 0. The Iwasaki improved gluon action was used with 
a clover quark action.
(Ali Khan, 2002, CP-PACS collaboration)}
\label{spectunq}
\end{figure}

\Figure{\ref{spectunq}} compares results for the masses of $\phi$ and
$K^{{\myast}}$ mesons on quenched and unquenched configurations as a function
of $a$. The $K$ meson is used to fix $m_s$ and gives poor results for
the $K^{\myast}$ and the $\phi$ in the quenched approximation, as described
earlier. For two flavors of dynamical quarks, the $K^{\myast}$ and $\phi$
masses are much closer to experiment, at least after a continuum
extrapolation (Ali Khan, 2002, CP-PACS collaboration).  One worrying
feature of this plot is the size of discretisation errors in the
unquenched case, implying that the improved action used does not work
as well in that case.

\begin{figure}
\centerline{\includegraphics[height=7.0cm]{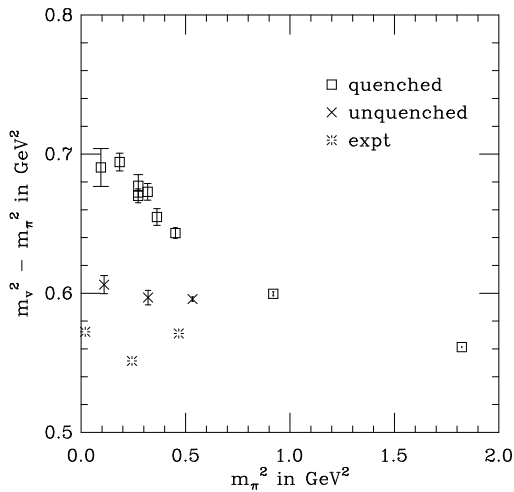}}
\caption{The difference of the squares of the vector and pseudoscalar
masses for various light hadrons, obtained with quenched and dynamical
lattice QCD. The dynamical results have 2+1 flavors of dynamical
quarks with masses $\approx m_s/4$ and $m_s$. The dynamical results
are given only for valence quark masses equal to the dynamical
ones. Experimental results are given by the bursts, using an estimated
mass for the pseudoscalar $s\overline{s}$ meson. The lattice spacing
has been obtained using $r_0 = 0.5$fm. Errors do not include errors
from fixing the lattice spacing (Bernard, 2001, MILC collaboration). }
\label{milcmvmpi}
\end{figure}

\Figure{\ref{milcmvmpi}} shows another quantity from light hadron
physics that gives a problem in the quenched approximation.  This is
the difference of the squared vector and pseudoscalar masses for given
quark combinations. Experimentally the result is very flat as a
function of quark mass, being $\approx 0.55 {\rm GeV}^2$ from the
$\pi, \rho$ to the $D, D^{\myast}$. In the quenched approximation this
quantity has a pronounced downward slope as the quark mass is
increased. Recent results from the MILC collaboration with 2 ($m_s/4$)
+ 1 ($m_s$) flavors of dynamical improved staggered quarks show
qualitatively different behaviour, much closer to that of experiment
(Bernard, 2001, MILC collaboration). This is the strongest indication
yet that calculations with dynamical quarks will overcome the
disagreements between the quenched approximation and experiment.

\section{Lattice QCD results}

The Proceedings of each year's lattice conference provide a useful
summary of current results and world averages. See (LAT2000, LAT2001).
Almost all lattice papers can be found on the hep-lat archive,
http://arXiv.org/hep-lat/. I have deliberately chosen to refer to
reviews where possible and these should be consulted for fuller access
to the literature.

%

%
\subsection{Methods for heavy quarks}

\index{heavy quark physics}Bottom and charm quarks are known as heavy
quarks since they have masses much greater than the typical QCD scale,
$\Lambda_{\rm QCD}$, of a few hundred MeV. Top quarks are also heavy,
of course, but do not have interesting bound states so are not studied
by lattice QCD. $b$~and $c$ quarks could be treated in the same way as
$u$, $d$, or $s$ quarks on the lattice except that, with current
lattice spacings of about 0.1fm, we have $m_ba$ in the interval 2--3
and $m_ca$ in 0.5--1. If $ma$ is not small then discretisation errors
of the form $ma$, $(ma)^2$ etc. will not be small either and such an
approach will not give accurate results.  Relativistic momenta,
$\null{p} \approx m$, can also not be well simulated if $\null{p}a$ is
not small: $\null{p}a$ of $O(1)$ corresponds to wavelengths which are
in danger of being small enough to `fall through' the holes in the
lattices.

To reach the very fine lattices that would be required to give $m_ba
\ll 1$ and accurate simulations for $b$ quarks would require an amount
of computing power way beyond our current hardware even in the
quenched approximation.  Luckily the physics of heavy quark systems in
the real world means that we do not have to do this; indeed, it would
be largely a waste of computer power. $b$ and $c$ quarks are
non-relativistic in their bound states, so that $m$ and $\null{p}
\approx m$ are irrelevant dynamical scales. The non-relativistic
nature is evident from the hadron spectrum.  There are heavy-heavy
bound states in which both the valence quark and anti-quark are heavy
($\Upsilon$, $\psi$ and $B_c$) and heavy-light bound states in which
the heavy (anti-)quark is bound to a light partner ($B$, $B_s$, $D$,
$D_s$) or partners, in the case of baryons ($\Lambda_b$,
$\Lambda_c$). In all cases the mass difference (splitting) between
excitations of these quark systems is much less than the mass of the
hadrons. For example \mbox{$m(\Upsilon^{\prime}) {-} m(\Upsilon)=560$MeV},
$m(\Upsilon)$ = 9.46GeV. The internal dynamics, which controls these
splittings, operates with scales much smaller than the quark
mass. Instead the important scales are the typical momentum carried by
the quark inside the bound state, $mv$, and the typical kinetic
energy, $\frac{1}{2}mv^2$. That these scales are small compared to $m$
implies that $v/c \ll 1$.  The use of non-relativistic techniques on
the lattice is then a good match to the physics of $b$ and $c$ systems
as well as providing an efficient way to handle them numerically on
the lattice.

There are several ways to proceed, and it is important when reading
the lattice literature to understand which method has been used.
In the remainder of this section we consider three methods in particular:
(a) static quarks, (b)  NRQCD (a non-relativistic version 
of QCD) and (c)  heavy relativistic quarks.

\subsubsection*{Static quarks} 

\index{static quarks}This is the $m =
\infty$ limit of heavy quarks.  In this limit Heavy Quark Symmetry
holds and quarks become static sources of colour charge with no spin
or flavor. This is evident on the lattice as the quark propagator
becomes simply a string of gluon fields along the time direction
(Eichten, 1990). Obviously no real quarks have infinite mass but this
is a useful limit for studying heavy-light systems. Corrections away
from the infinite mass limit are the subject of Heavy Quark Effective
Theory (Buchalla, 2002).

\subsubsection*{NRQCD} 

NRQCD is a non-relativistic version of QCD (Lepage,
1992). \index{non-relativistic QCD}The Lagrangian for heavy quarks is
the non-relativistic expansion of the Dirac Lagrangian:
\begin{equation}
{\cal{L}}_Q = \overline{\psi} ( D_t - \frac {\mathbf{D}^2} {2m_Qa} - c
\frac {\bm{\sigma}{\cdot} \mathbf{B}} {2m_Qa} + \cdots ) \psi 
\label{lnrqcd}
\end{equation}
where additional terms can be added to go to higher order in
$v/c$. $\psi$ is now a 2-component spinor since the quark and
anti-quark fields of the Dirac fields decouple from each other. $D$ is
a covariant derivative, including coupling to the gluon
field. $\mathbf{B}$ is the chromomagnetic field, related to space-space
components of the field strength tensor, $B_i =
\epsilon_{ijk}F_{jk}$. $m_Q$ is the quark mass; heavy quarks are
frequently generically denoted $Q$ in contrast to the $q$ used for
light quarks. Notice that the quark mass term
$\overline{\psi}m_Qa\psi$ has been dropped. This simply redefines the
zero of energy so that the energies of all hadrons in lattice units
are less than 1.

The NRQCD Lagrangian can be discretised onto a lattice and leads to
much simpler and faster numerical algorithms for calculating the quark
propagator than for light quarks. Instead of having to explicitly
invert a matrix using an expensive iterative procedure such as
Conjugate Gradient, the propagator is simply calculated by stepping
through the lattice in time and calculating the propagator at time $t$
from that at time $t-1$.  This is simply illustrated if we look at the
Lagrangian in the infinite mass limit, where it becomes the Lagrangian
for static quarks. Only the first term above contributes and we have:
\begin{equation}
S_Q = \sum_x \overline{\psi}(x)
\left(\rule[-1ex]{0mm}{3ex}U_t(x)\psi(x+1_t)-\psi(x)\right)
\end{equation}
$M$ is then an upper triangular matrix, 
using the notation of Equation~\ref{incq},
and the quark propagator is given by:
\begin{equation}
(M^{-1,Q}_{0,t+1}) = U^{\dag}_t (M^{-1,Q}_{0,t}).
\end{equation}
The general start and end points, $x$ and $y$, are simply denoted here
by their $t$ co-ordinates, 0 for the origin and $t$ for the end
point. To move from end point $t$ to $t+1$ just requires
multiplication by the appropriate $U$ field in the time direction, so
$M^{-1}$ does not change spatially and becomes a string of $U$ fields
as described for static quarks above.  For NRQCD with non-infinite
masses, the evolution equation in $t$ for the propagator is not as
simple and does contain spatial variations (e.g. from the spatial
covariant derivatives in Equation~\ref{lnrqcd}) but the same
principles apply. A smearing function, $\phi$, is chosen at the time
origin and then the propagator calculated from 0 to later times by an
evolution equation from one $t$ to the next.  This makes NRQCD
numerically very attractive. Heavy quark propagators, once calculated,
can be combined together or with a light quark propagator to make 2-
and 3-point functions for heavy hadrons as described for light hadrons
earlier. As described there also, the value for the bare heavy quark
mass in lattice units, $m_Qa$, is adjusted, given a value for $a$,
until a heavy hadron mass is correct in GeV. The energies of heavy
hadrons calculated on the lattice do not in fact equate directly to
their masses because the mass term was removed from the
Lagrangian. Instead, for one heavy hadron we have to calculate an
energy-momentum dispersion relation and derive its mass from the
momentum dependence ($E \propto \mathbf{p}^{\, 2}/2M$).

NRQCD is an effective theory, containing the right physics for low
momentum heavy quarks. Adding more relativistic corrections to the
Lagrangian can make this more accurate. These higher order terms
appear with coefficients (such as $c$ in equation~\ref{lnrqcd}) which
must be determined by matching to relativistic QCD. These coefficients
represent the effect of relativistic momenta missing from NRQCD and
they are governed by $\alpha_s$ at this high momentum scale and so are
perturbative. High momenta for both quarks and gluons are missing
anyway on the lattice because of the discretisation of space-time. We
described earlier how a better match between lattice QCD and QCD is
made by adding terms to the lattice QCD Lagrangian which are higher
order in $a$, with a coefficient which depends on the strong coupling
constant at the lattice cut-off scale. That the two procedures are
very similar is not an accident; indeed, the same higher dimension
operators appear in both cases. In this case NRQCD is simply making a
virtue of the existence of the lattice cut-off. The difference is,
however, that in the NRQCD case the operators appear with inverse
powers of $m_Qa$ (in a dimensionless lattice notation) and so $m_Qa$,
and therefore $a$, cannot be taken to zero in this approach. NRQCD has
no continuum limit, but this does not prevent physical results being
obtained at finite lattice spacing.  It is just necessary to show that
the results are sufficiently independent of $a$ over a range of values
of $a$.

\subsubsection*{ Heavy relativistic quarks} 

This method looks very
different from NRQCD, but has a lot of features in
common. \index{heavy relativistic quarks}The use of a relativistic
action, such as the Wilson/clover action, for heavy quarks on a
lattice does not have to be incorrect if the results are interpreted
carefully (El-Khadra, 1997).  The main point to realise is that the
existence of a large value for $ma$ breaks the symmetry between space
and time. The inverse quark propagator in momentum space has an energy
at zero momentum very different from its mass (e.g. for a free Wilson
quark, $E(\mathbf{p}=0) = \ln(1+ma)$) but its momentum dependence for
small momenta is correct (i.e.  as $\mathbf{p}^{\, 2}/ma$). Thus, we can
ignore the $ma$ errors in the energy if we fix masses from the
energy-momentum relation as for NRQCD.  For more precision we must add
higher order discretisation/relativistic corrections. These will
appear with coefficients chosen to match continuum relativistic
QCD. As we have seen the coefficients are a power series in $\alpha_s$
at the cut-off scale and they will depend on $ma$.  For small $ma$ the
coefficients will be those of a discretisation correction to the
action; for large $ma$ they will go over to the NRQCD
coefficients. For example, the $\sigma_{\mu\nu}F_{\mu\nu}$ clover term
corrects for an ${O}$$(a)$ error in the Wilson action for light
quarks; for heavy quarks, it becomes the relativistic correction which
couples the quark spin and the chromomagnetic field.  In this way an
action can be developed that smoothly interpolates between heavy and
light quark physics, at the numerical cost of having to handle heavy
quarks in the same way as light ones. This method is sometimes known
as the Fermilab method, since it was pioneered there.

The charm quark mass is not very heavy on the finest of current
quenched lattices, and some groups have taken the standard
relativistic approach in this case. To reach the $b$ quarks then
requires an extrapolation jointly in the heavy quark mass and the
lattice spacing (Maynard, 2002, UKQCD collaboration) to avoid
confusing discretisation and relativistic corrections. Such an
extrapolation inevitably has rather large errors.  A better approach
is to consider a formalism which explicitly breaks space-time symmetry
in order to restore the relativistic energy-momentum relation for
heavy quarks. For example, you can take an anisotropic lattice which
has a much finer spacing in the time direction than in the space
directions. $ma_t$ is then small and the heavy quark looks like a
light one, at the cost of having many more timeslices on the lattice,
and having to determine the lattice spacing in both directions (Chen,
2001).

\subsection{The heavy hadron spectrum}

\begin{figure}
\rule{40mm}{0mm}{(a)}\rule{70mm}{0mm}(b)\\[1ex]
\centerline{
\includegraphics[width=75mm,clip,trim=0 130 10 0]{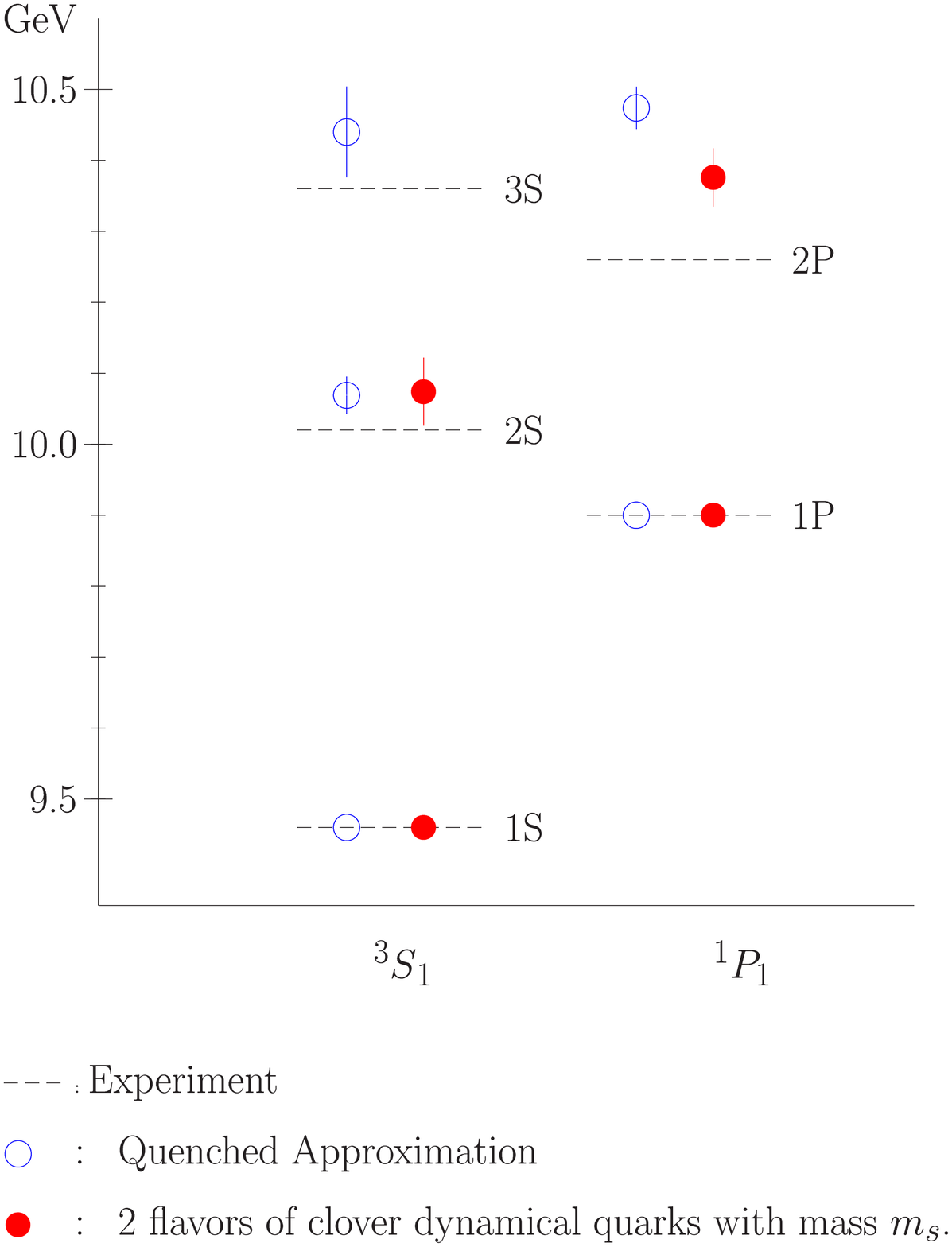}
\includegraphics[width=90mm,height=80mm,clip,trim=0 145 10 0]{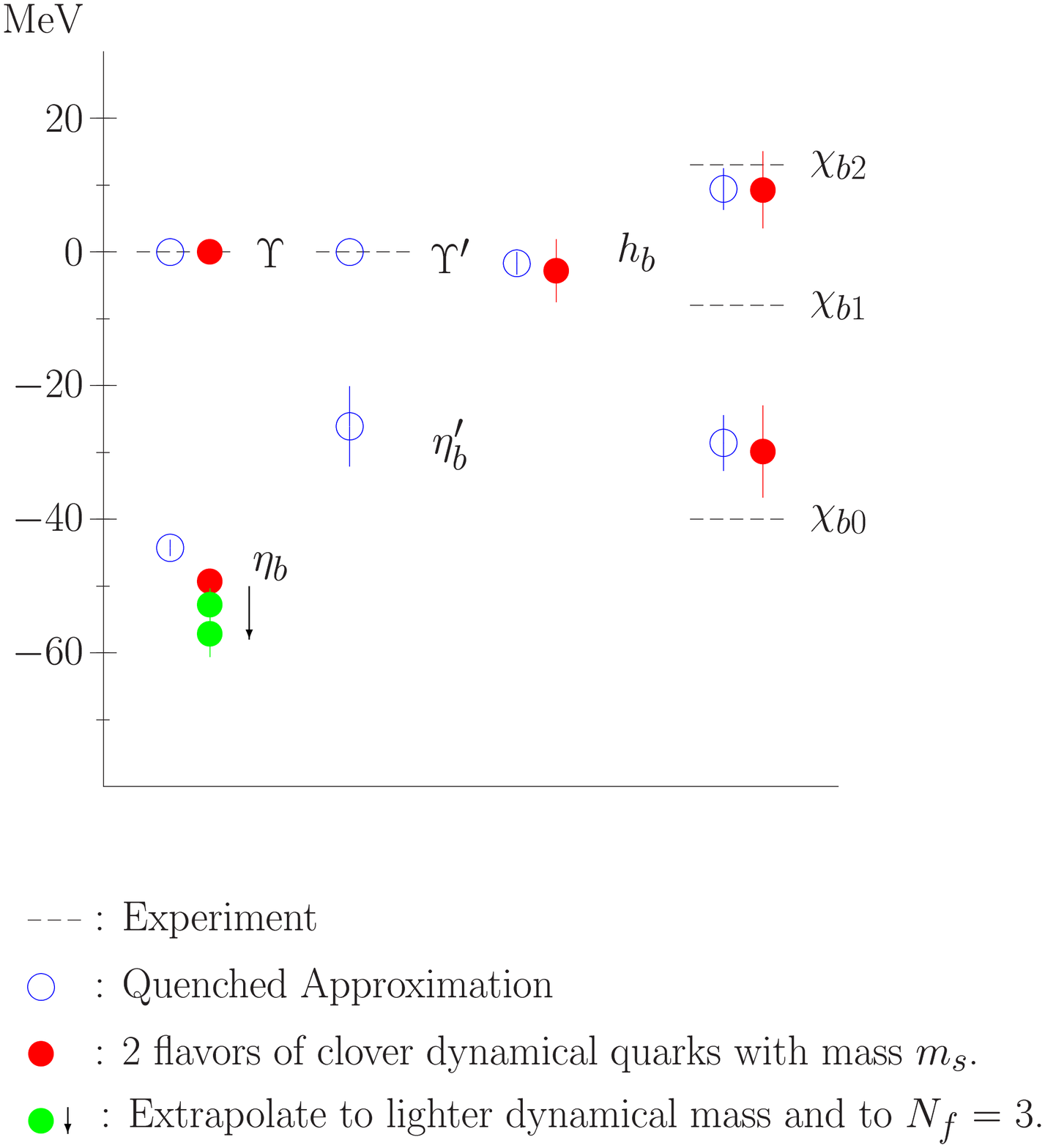}
}
\caption{(a) The radial and orbital excitations in the $b\overline{b}$
system, as calculated in lattice QCD using NRQCD for the $b$ quarks
(Marcantonio, 2001, UKQCD collaboration).
(b) The fine structure of low-lying $b\overline{b}$ states.
Legend: the horizontal dashed lines are experimental values; open
circles show the quenched approximation; solid circles correspond to 2
flavors of clover dynamical quarks with mass $m_s$. (The lowest cluster
of points on the right show an extrapolation to lighter dynamical mass
and to $N_f{=}3$.)
}
\label{upssi}
\end{figure}

\index{heavy hadron spectrum}The spectrum of heavy-heavy states has
largely been the province of NRQCD (Davies, 1998).  \Figure{\ref{upssi}(a)}
shows the radial and orbital excitations of the $b\overline{b}$
$\Upsilon$ system, \index{bottomonium}obtained both on quenched gluon
configurations and those with two flavors of dynamical quarks
(Marcantonio, 2001, UKQCD collaboration). For these results the
lattice spacing has been fixed by demanding that the splitting between
the $\Upsilon(1S)$ and the spin-average of the $P$-wave ($\mychi_b$)
states is correct. The $b$ quark mass has been fixed by requiring that
the $\Upsilon(1S)$ mass be correct. It is only the $2S$
($\Upsilon^{\prime}$), $3S$ ($\Upsilon^{\prime \prime}$) and $2P$
($\mychi_b^{\prime}$) states that are predicted from this calculation,
and they have rather large statistical errors at present. It is a
general feature of lattice calculations that ground state masses are
more precise than excited state masses. For both excited and ground
states the noise is controlled by the ground state mass. For excited
states the signal/noise ratio is then much worse and becomes
exponentially bad at large $T$.

Of more immediate interest is the fine structure of the low-lying $S$
and $P$ states, shown in \Figure{\ref{upssi}(b)}.  These can be determined
very precisely on the lattice, particularly the `hyperfine' splitting
between the spin-parallel vector $\Upsilon$ state and the not-yet-seen
spin-antiparallel pseudoscalar $\eta_b$. A comparison with experiment,
when it exists, for this splitting will provide a very good test of
lattice QCD and our $b$ quark action, which will be important for the
lattice predictions of $B$ matrix elements described in Section 4.3.

The accuracy of the NRQCD, or other lattice action, for heavy-heavy
bound states can be estimated by working out what order in an
expansion in powers of $v/c$ is represented by each term. e.g. the
first two terms in the NRQCD action of Equation~\ref{lnrqcd}, i.e. the
time derivative and the kinetic energy term, are both
${O}$$(v^2/c^2)$. This is because the `potential energy' and
kinetic energy terms are roughly equal for two heavy particles. These
terms give rise to the radial and orbital splittings, and the ratio of
these ($\approx$ 500MeV) to half the $\Upsilon$ mass gives an
estimate of $v^2/c^2 \approx 0.1$ for $b$ quarks in an $\Upsilon$.
Higher relativistic corrections, such as the $\mathbf{D}^4/8m_Q^3$ term,
are ${O}$$(v^4/c^4)$ and should give roughly a 10\% correction to
these splittings. These terms were included here, but not the
$v^6/c^6$ corrections, so an error of roughly 1\% remains. The
$\bm{\sigma}\cdot\mathbf{B}$ term of Equation~\ref{lnrqcd} is the first
spin-dependent term and is ${O}$$(v^4/c^4)$.  It gives rise to the
hyperfine splitting and a similar term of the same order, proportional
to $\bm{\sigma}\cdot\mathbf{D}\times\mathbf{E}$, gives rise to the $P$ fine
structure. The fine structure is indeed roughly 10\% of the radial and
orbital splittings. Including only these terms in the NRQCD action, as
was done here, implies an error of roughly 10\% in these splittings.
A more precise calculation, necessary to test this action against
experiment, will require the $v^6/c^6$ spin-dependent terms and the
$\alpha_sv^4/c^4$ terms implied by calculating the coefficient $c$ in
front of the $\bm{\sigma}\cdot \mathbf{B}$ term in
equation~\ref{lnrqcd}. This is now being done.  Figure~\ref{upssi}(b)
does show, however, that the hyperfine splitting increases when two
flavors of dynamical quarks are included, and continues to increase as
the dynamical quark mass is reduced towards real $u$ and $d$ quark
masses. We expect the $\Upsilon$ to see also $s$ quarks in the vacuum
and extrapolating the number of dynamical flavors to three increases
the splitting further.

\begin{figure}
\centerline{\includegraphics[height=7.0cm]{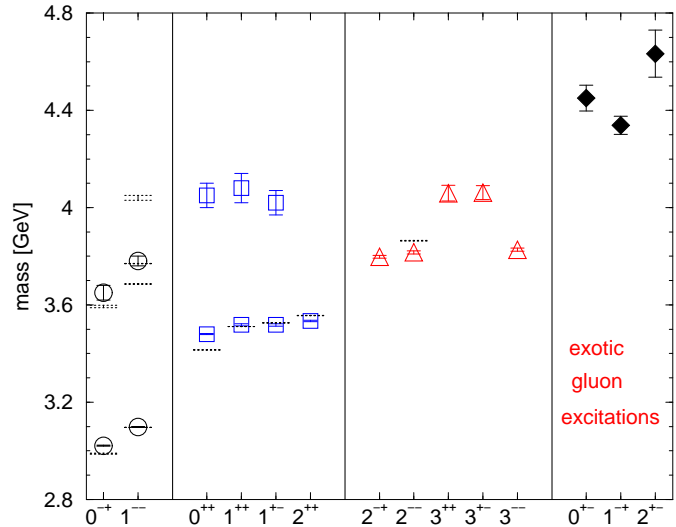}}
\caption{The spectrum of $c\overline{c}$ states, as calculated in
lattice QCD using anisotropic quenched configurations (Chen, 2001).}
\label{ccbar}
\end{figure}

The charmonium\index{charmonium}, $\psi$, system is more relativistic
that the $\Upsilon$ system and correspondingly less well-suited to
NRQCD.  Estimates as above give $v^2/c^2 \approx
0.3$. \Figure{\ref{ccbar}} shows the charmonium spectrum obtained from
anisotropic relativistic clover quarks in the quenched approximation
(Chen, 2001).  The lattice spacing and charm quark mass were fixed in
the analogous way to that described above, except that the spin
average of the vector $J\psi$ and pseudoscalar $\eta_c$ masses was
used to fix $m_c$. Since the $\eta_c$ mass is known experimentally
this gives improved precision since the spin-average is not sensitive
to any inaccuracies in spin-dependent terms.  The spectrum given in
Figure~\ref{ccbar} includes some gluonic excitations of the
$c\overline{c}$ system, i.e. $c\overline{c}g$ states, called
hybrids. Their existence is expected simply from the non-Abelian
nature of QCD which allows gluons themselves to carry color
charge. Some of these hadrons have exotic quantum numbers not
available to mesons made purely of valence quarks, and the prediction
of their masses will be important for their experimental discovery.

\begin{figure}[t]
\centerline{\includegraphics[height=8.0cm]{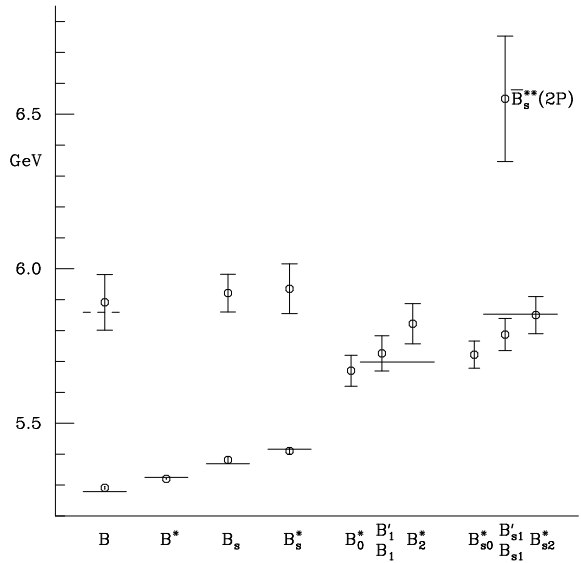}}
\caption{The spectrum of bound states of a $b$ quark with 
a light anti-quark as
calculated in lattice QCD in the quenched approximation using NRQCD 
for the $b$ quark (Hein, 2000).}
\label{bspect}
\end{figure}

\Figure{\ref{bspect}} shows the spectrum of mesons made from one $b$
quark and one light ($u/d$ or $s$) anti-quark in the quenched
approximation\index{B spectrum} (Hein, 2000).  NRQCD was used for the
$b$ quark, and the clover action for the light quark. In this case the
lattice spacing was fixed using a quantity from the light hadron
spectrum, $m_{\rho}$, because heavy-light systems are more similar in
terms of internal momentum scales to light hadrons than heavy-heavy
ones. See the comments in Section~3.2 on how the lattice spacing in
the quenched approximation depends on the quantity used to fix it. The
$u/d$ and $s$ quark masses were fixed using the $\pi$ and $K$ masses.
The $b$ quark mass was fixed from the spin-average of the $B$ and
$B^{\myast}$ meson masses. Taking a spin-average, as above for charmonium,
avoids any errors from spin-dependent terms in the action. The $b$
quark mass obtained this way differs from that obtained above from the
$\Upsilon$ system, and is another feature of the quenched
approximation. In the `real world' there is only one lattice spacing
and one set of quark masses and parameters fixed from the $\Upsilon$
system will be used to predict the entire $B$ spectrum.

The power counting in $v/c$ for terms in the Lagrangian works rather
differently in heavy-light systems compared to heavy-heavy ones.  Now
there is one quark that carries almost all the mass of the heavy-light
system and it sits in the centre surrounded by the swirling light
quark cloud. This picture makes sense even in the limit in which the
heavy quark has infinite mass when the Lagrangian would contain only
the covariant temporal derivative $D_t$ (static quarks). The higher
order terms in the Lagrangian can then be ordered in terms of the
inverse powers of the heavy quark mass that they contain. This is
equivalent to an expansion in powers of $v/c$. The typical momentum of
a heavy quark in a heavy-light system is ${O}$$(\Lambda_{\rm QCD})$
(as is that of the light quark) and so $v/c \approx
\Lambda_{\rm QCD}/m_Q$. This gives $v/c \approx$ 10\% for the $B$ and 30\%
for the $D$.

Again the power counting exercise enables us to understand the
approximate relative sizes of different mass splittings in the
spectrum and the accuracy of our lattice QCD calculation to a given
order in $v/c$. The leading spin-independent term in the action is
$D_t$ giving rise to the orbital and radial excitations of $\approx$
500MeV. The kinetic energy term, $\mathbf{D}^2/2m_Q$ gives a
$\Lambda_{\rm QCD}/m_Q$ correction to this, which depends on the quark
mass and, therefore flavor. This explains why these excitation
energies are so similar for $B$ and $D$ systems; the similarity
between $\psi$ and $\Upsilon$ is more accidental.  The leading
spin-dependent term is $\bm{\sigma}{\cdot} \mathbf{B}/2m_Q$, which gives
rise to fine structure such as the splitting between the pseudoscalar
$B$ and vector $B^{\myast}$. This splitting should then be smaller by a
factor of $\Lambda_{\rm QCD}/m_Q$ compared to the spin-independent
splittings and this is indeed observed.  To calculate this splitting
precisely on the lattice requires the inclusion of higher order terms
in the Lagrangian, as well as a better matched coefficient $c$ for the
$\bm{\sigma}\cdot \mathbf{B}$ term and this will be done in future
calculations.

We have stressed that lattice QCD is simply a way of handling QCD. It
has the same a priori unknown parameters as QCD, the overall scale
(equivalent to the coupling constant) and the quark masses. These
parameters come from a deeper theory and must simply be fixed in the
QCD Lagrangian using experiment and the results from a calculation in
QCD.  As described in Section 3, Lattice QCD provides the most direct
way of doing this.  The values for the parameters obtained are then
useful input to other theoretical techniques.

\begin{figure}
\centerline{\includegraphics[height=9.0cm]{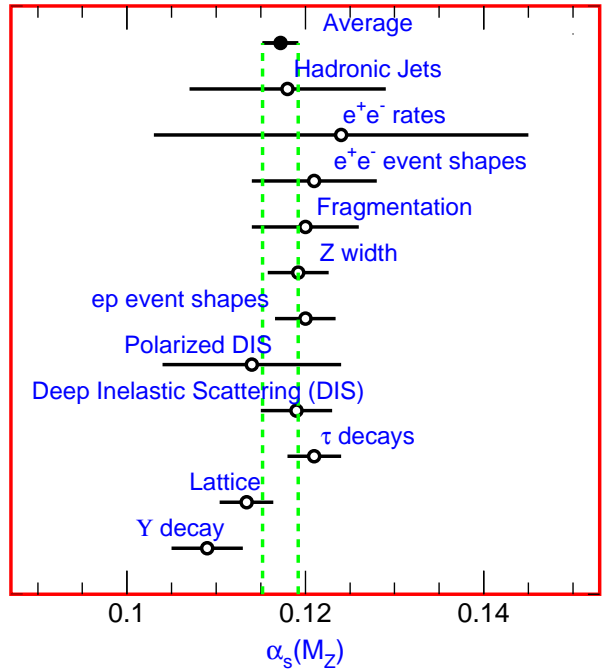}}
\caption{A comparison of determinations of the strong 
coupling constant, expressed as $\alpha_s^{\overline{MS}}(M_Z)$ (PDG, 2001).}
\label{alphacomp}
\end{figure}

Determination of the lattice spacing at a given lattice bare coupling
constant, is equivalent to (and can be converted into) a determination
of the renormalised coupling constant, $\alpha_s$, at a physical scale
in GeV. To compare to other determinations of $\alpha_s$, this can be
converted to the $\overline{MS}$ scheme and run to
$M_Z$. \Figure{\ref{alphacomp}} shows a comparison of different
determinations of $\alpha_s$ from the Particle Data Group (PDG, 2001).
It is clear that the lattice result is one of the most
precise\index{alphas@$\alpha_s$}.

All methods for determining $\alpha_s$ have three components:
\begin{enumerate}
\item
Theoretical input: a perturbative expansion 
in $\alpha_s$, for some quantity.
\item
A value for that quantity.
\item
An energy scale.
\end{enumerate}
Most methods use an experimental result for stage 2, where the lattice 
uses a non-perturbative evaluation on the lattice of 
the vacuum expectation value of a simple 
short-distance gluonic operator (such as the plaquette). This avoids 
the problems of hadronisation etc which reduce the precision of 
methods based on the experimental determination of jet shapes
or cross-sections. All methods use experiment for stage 3, and 
here the lattice-based determination needs an experimental result 
to fix the lattice spacing. A good quantity to use here is 
the orbital excitation energy ($1P{-}1S$) in, say, the $\Upsilon$ system since 
this is well-determined on the lattice and directly measured experimentally.

Quark masses are also well determined on the lattice.  Since quarks
are not freely available to be weighed, as an electron would be, care
must be taken in defining what exactly is meant by the quark mass.
The bare mass in the lattice QCD Lagrangian for a particular action,
determined by the requirement to get a particular hadron mass correct
and converted to physical units, is a well-defined quantity but not
very convenient.  We can convert it perturbatively into, say, the
running quark mass in the $\overline{MS}$ scheme.  The best current
determination of the $b$ quark mass is in fact from the static
approximation in which $b$ quarks have infinite mass. There is no bare
$b$ quark mass in that case; instead the binding energy $m_B - m_b$ is
calculated, and from that, $m_b$ is determined. The binding energy is
small compared to $m_B$ and has only weak dependence on the $b$ quark
mass, so for this quantity the static approximation is a good one. The
$b$ quark mass \index{mb@$m_b$}obtained in this way is 4.30(10)GeV in
the quenched approximation, with some indications that it is slightly
lighter when dynamical quarks are included (Lubicz, 2001).

\begin{figure}[!b]
\centerline{\includegraphics[height=7.0cm]{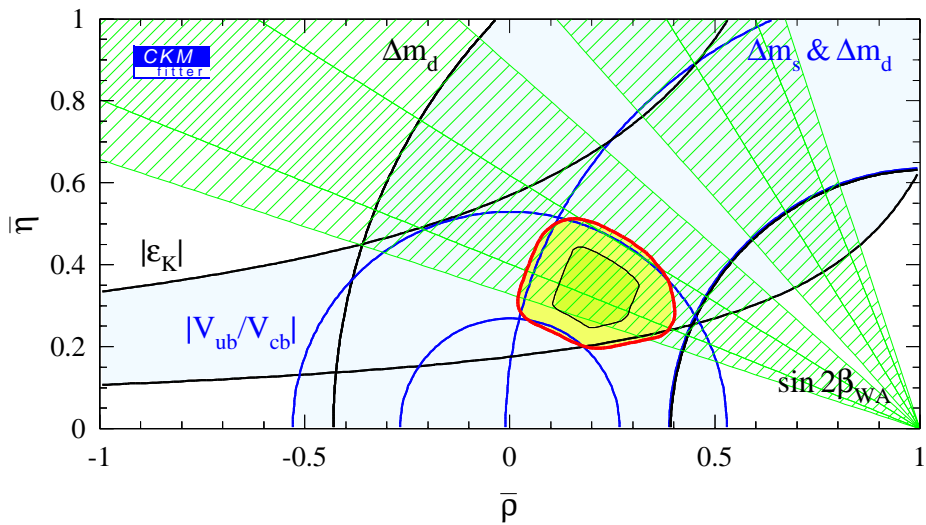}}
\caption{The unitarity triangle with constraints on the 
upper vertex  obtained from different quantities (Hocker, 2001). 
The lower vertices are at $\overline{\eta}$ = 0, $\overline{\rho}$ = 0 
and 1.} 
\label{unitri}
\end{figure}

\subsection{Heavy hadron matrix elements}

Precise lattice calculations of matrix elements for $B$ decay are
essential to the experimental $B$ factory programme (Stone, 2002).
This aims to test the internal consistency of the Standard Model in
which CP violation occurs through the Cabibbi-Kobayashi-Maskawa
matrix. The weak decays of the $b$ quark are particularly useful in
giving us access to poorly known elements of this matrix.  The
unitarity of the CKM matrix can be represented by a triangle; the
position of the upper vertex being constrained by the determination of
angles and sides, see \Figure{\ref{unitri}}.  The angles are determined
directly by measurement of asymmetries. The determination of the sides
requires both the experimental measurement of a decay rate and its
theoretical calculation.This allows the magnitude of one of the CKM
elements to be extracted. Below we describe the lattice calculation of
the matrix elements most important for this programme. The extent to
which the unitarity triangle can be tested depends on both the
experimental and the theoretical errors. It is critical to reduce the
errors from lattice calculations to a few percent, otherwise they will
dominate the uncertainties from experiment.

The simplest 2-point matrix element that can be calculated on the
lattice is that for the decay constant of the charged pseudoscalar
heavy-light mesons (see Figure~\ref{2pt})\index{B decay constant}.
For the $B$ this is known as $f_B$ and it is obtained from the vacuum
to $B$ matrix element of the axial vector current which couples to the
$W$.
\begin{equation}
\langle 0 | A_{\mu} | B \rangle = p_{\mu} f_B.
\label{fB}
\end{equation}
The purely leptonic decay rate of the $B$ meson is then proportional
to $f_B^2$ times kinematic factors times the square of the CKM element
which multiplies the appropriate axial vector current in the
Lagrangian, in this case $\overline{u}\gamma_{\mu}\gamma_5 b$ (Rosner,
2002).  In principle an experimental determination of the leptonic
decay rate could be combined with the lattice calculation to yield
$V_{ub}$, but in practice the experiment is very hard to do because
the rate is so low.  For other heavy-light mesons, it may be possible.
$f_{D_s}$ has been measured experimentally, but not very precisely as
yet. It can be used, with lattice calculations, to give  $V_{cs}$.

It is important to realise that, although we are discussing the weak
decay of a $b$ or $c$ quark, the calculations are done in lattice QCD.
The quark cannot decay in isolation, but must be bound into a hadron
by the confinement property of QCD. The determination of the decay
matrix element must then take into account all the QCD interactions
inside the hadron (see Figure~\ref{2pt}) and this requires lattice
QCD. We do not put the $W$ boson on the lattice.  As far as QCD is
concerned the $B$ meson annihilates into the vacuum.  The virtual $W$
boson decay to leptons is put in by hand when we calculate the decay
rate.

\begin{figure}[!b]
\centerline{\includegraphics[height=6.0cm]{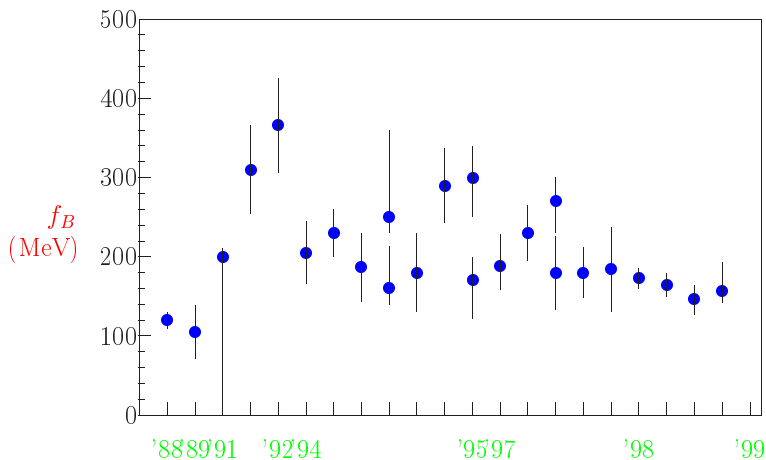}}
\caption{A timeline of results for the $B$ meson decay constant,
$f_B$, calculated in lattice QCD in the quenched approximation. }
\label{timeline}
\end{figure}

Lattice calculations of $f_B$ improved markedly through the 1990s
(this has been true of most lattice calculations) as we got to grips
with the systematic errors. \Figure{\ref{timeline}} shows a timeline
of results in the quenched approximation. It shows both that lattice
calculations have markedly improved and that early calculations had
very unreliable estimates of their errors.  The most recent and best
calculations do a careful job of matching the lattice representation
of the axial vector current to the continuum. For heavy-light mesons
we have to be careful both about relativistic ($\Lambda_{\rm
QCD}/m_Q$) corrections and discretisation corrections to the leading
order lattice current. Since $m_Qa$ is a dimensionless number, these
two corrections in fact appear together and can be considered
simultaneously. The matching between lattice and continuum is
currently done only to ${O}$$(\alpha_s)$ and this is the major source
of error in the quenched approximation.  \Table{\ref{errors}} shows a
typical `error budget' for such a calculation.  We need a more precise
matching, either to $\alpha_s^2$ or non-perturbatively (both of which
can be done with a lot of hard work), to improve the errors beyond the
10\% level.

\begin{table}[!h]
\centering
\begin{tabular}{|l|c|}\hline
Source &  percent error  \\\hline
statistical + fitting &  $3$  \\
discretisation $O((a\Lambda)^2)$ &    4   \\
perturbative $O(\alpha_s^2,\alpha_s^2/(aM))$ & 7  \\
NRQCD $O((\Lambda/M)^2,\alpha_s\Lambda/M)$ &  2\\
light quark mass  &  4  \\
$a^{-1}(m_\rho)$&  4   \\\hline
Total &  {\bf 10}\\\hline
\end{tabular}
\caption{Source of error in a typical lattice calculation of $f_B$ 
using NRQCD for the heavy quark in the quenched approximation. $a \approx 
0.1$fm, $M$ is the $b$ quark mass, $\Lambda$ a typical QCD scale 
of a few hundred MeV and $\alpha_s$ is evaluated at $2/a$.}
\label{errors}
\end{table}

Recent reviews of lattice results (Ryan, 2002), (Bernard, 2001) have
given the following `world averages' for lattice results in the
quenched approximation:
\begin{itemize}
\vspace{-1ex}
\item  $f_B$ = $173 \pm 23$MeV 
\vspace{-1ex}
\item  $f_{D_s}$ = $230\pm 14$MeV 
\vspace{-1ex}
\item  $f_{B_s}/f_B$ = 1.15(3); $f_{D_s}/f_D$ = 1.12(2).
\end{itemize}
(Note that the $B_s$ does not decay purely leptonically but the
calculation of the appropriate matrix element can still be done in
lattice QCD and yields useful information on its dependence on the
light quark mass.)  Large-scale calculations on dynamical
configurations are only just beginning, so unquenched results are
still unclear. It seems likely that decay constants will be 10--20\%
larger unquenched.

\begin{figure}
\centerline{\includegraphics[height=30mm]{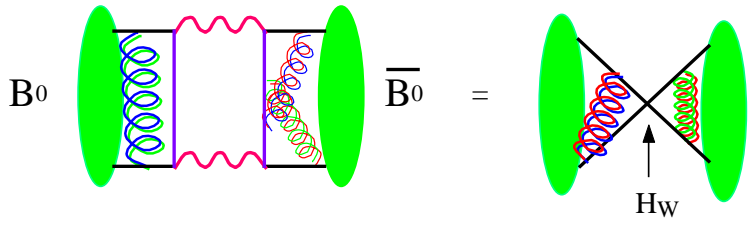}}
\caption{ The $B$ box diagram, related to that of a 4-quark operator. }
\label{box}
\end{figure}

A more important quantity from the point of view of the $B$ factory
programme is the mixing amplitude for neutral $B$ mesons, $B^0$ and
$B_s$. \index{B mixing}This mixing gives rise to a difference in mass
between the CP-eigenstates, $\Delta m$, which can be measured
experimentally through oscillations between particle and anti-particle
(Stone, 2002).  The mixing amplitude is given by the `box diagram'
(see \Figure{\ref{box}}) in which the $b$ quark and light anti-quark
convert to a $b$ anti-quark and light quark through the mediation of
virtual $W$s and (preferentially) $t$ quarks. The mixing amplitude is
then proportional to the matrix element of the box between, say, a
$B^0$ and a $\overline{B}^0$ multiplied by the product of CKM elements
$V_{tb}^{\myast}V_{td}$. The current determination of $|V_{td}|^2$ from
experiment and theory gives a curve on the unitarity triangle plot
(marked $\Delta m_d$ on Figure~\ref{unitri}).  Future experiments will
be able to see oscillations of the $B_s$ and then ratios of $\Delta
m_{B_s}/\Delta m_{B}$ will allow a more precise determination of
$|V_{ts}/V_{td}|^2$, since some of the systematic errors will cancel
out.

As explained earlier, $W$ bosons do not appear in lattice QCD
calculations.  The matrix element of the box diagram is calculated in
lattice QCD by replacing it with the equivalent 4-quark operator which
appears in the effective (low-energy) weak Hamiltonian (Rosner, 2002).
Conventionally this matrix element for the $B$ is set equal to $(8/3)
f_B^2 M_B^2 B_B$, giving a definition of the parameter confusingly
called $B_B$.  $B_B$ is the amount by which the matrix element differs
from the result that would be obtained by saturating the $H_W$ vertex
of Figure~\ref{box} with the vacuum (comparing this to the right hand
picture of Figure~\ref{2pt} we can see that this would be
$f_B^2$). $B_B$ is generally expected to be roughly 1, and this
explains why lattice calculations originally concentrated on
calculating $f_B$. To calculate $B_B$ is harder, but is now being
done. It requires, as for $f_B$, a careful matching between the
lattice and the continuum, and this has again been done to
${O}$$(\alpha_s)$ so far.

Recent world averages for the renormalisation-group-invariant definition 
of $B_B$ in the quenched approximation 
have been given as (Ryan, 2002), (Bernard, 2001):
\begin{itemize}
\vspace{-1ex}\item
$\hat{B}_{B_d}$ = 1.30(12)(13)
\vspace{-1ex}\item
 $f_{B_d}\sqrt{\hat{B}_{B_d}}$ =230(40)MeV
\vspace{-1ex}\item 
$\hat{B}_{B_s}/\hat{B}_{B_d}$ = 1.00(4) 
\vspace{-1ex}\item
$f_{B_s}\sqrt{\hat{B}_{B_s}}/f_{B_d}\sqrt{\hat{B}_{B_d}}$ = 1.15(6) 
\end{itemize}
A lot of the matching errors cancel out in the ratios between 
$B_s$ and $B_d$ so that the errors in these ratios are less than 
10\%. The ratio may also not be significantly affected 
by unquenching. 

Heavy-light mesons decay semi-leptonically \index{B semi-leptonic
decay}through a diagram in which the heavy quark changes flavor,
emitting a virtual $W$, and the other (spectator) quark in the meson
combines with the new quark flavor to make a new meson.  In this way
$B$ mesons can decay to $D$ or $D^{\myast}$ mesons if $b \rightarrow c$ and
to $\pi$ or $\rho$ mesons if $b \rightarrow u$. In each case the
appropriate CKM element appears at the current vertex in the
three-point diagram (see Figure~\ref{3pt}) and can therefore be
determined by a comparison of the experimental exclusive rate to the
theoretical one.  The ratio $V_{ub}/V_{cb}$ gives an important
circular constraint in the unitarity triangle (see
Figure~\ref{unitri}).

The calculation of the matrix element for $B$ semi-leptonic decay on
the lattice requires the calculation and simultaneous fitting of the
3-point function of Figure~\ref{3pt} and the appropriate 2-point
functions necessary to isolate the matrix element. It is therefore
significantly harder than a simple 2-point calculation. In addition
the matrix element depends on $q^2$, the squared difference of
4-momenta between the initial and final meson. This can take a range
of values, because the decay is a three-body one. The matrix element
can then be written as a combination of form-factors which are $q^2$
dependent, in contrast to the two-body leptonic decay which was
parameterised by a single number, $f_B$.  For example the pseudoscalar
to pseudoscalar transition (e.g. $B$ to $D$ or $\pi$) proceeds only
through the vector current and has two form factors, $f_{+}$ and
$f_0$:
\begin{equation}
\langle P^{\prime}(p^{\prime}) | V_{\mu} | P(p) \rangle = f_{+}(q^2)\left[ (p+p^{\prime})_{\mu} - \frac{M_P^2-M_{P^{\prime}}^2}{q^2}q_{\mu}\right] + f_{0}(q^2) \frac{M_P^2-M_{P^{\prime}}^2}{q^2}q_{\mu}
\label{formfs}
\end{equation}
The differential decay rate is proportional to the square of $f_{+}$
because the leptonic current $L_{\mu}$ coupling to the $W$ has 
$q^{\mu}L_{\mu} = 0$ for massless leptons. 
The pseudoscalar to vector transition proceeds through both the 
vector and axial currents and has 5 form factors, 3 of which appear in 
the decay rate. 

To explore different values of $q^2$ for semi-leptonic decay on the
lattice it is easiest to insert different 3-momenta at the final meson
and at the current, and then work out the resulting 3-momentum of the
initial meson. We are restricted to values of 3-momentum allowed on
the lattice, i.e. the components of $\mathbf{p}$ have the form $p_ja =
n_j 2 \pi/L$, where $n_j = 0,1,2 \ldots$ and $L$ is the number of
lattice sites in the $j$ direction.  The smallest non-zero value of
$p_j$ is then $2\pi/(La)$ where $La$ is the physical size of the
lattice in a spatial direction. A big physical volume is then required
to achieve a fine discretisation of momentum space and avoid a large
jump from one momentum to the next.  In general results at higher
momenta are much noisier than those at small momenta (this is for the
same reason that excited states are noisier than ground states,
discussed above) and calculations tend to be restricted to a few of
the smallest possible momenta. Discretisation errors will also be
larger at larger values of $pa$, so systematic errors will be higher.

For the matrix element for $B$ to $D^{({\myast})}$ semi-leptonic decay it is
useful to consider both the $b$ quark and the $c$ quark in the heavy
quark limit. In that limit, as discussed above, the Lagrangian for
heavy quarks becomes insensitive to the heavy quark spin or flavour
(Buchalla, 2002). The light quark cloud in the meson cannot tell
whether it is surrounding a $b$ or a $c$ quark or one whose spin is
pointing parallel or anti-parallel to its spin. Thus the form factors
for $B \rightarrow D$ and $B \rightarrow D^{\myast}$ will become identical
(or vanish) and the same as the $B \rightarrow B$ elastic form factor,
provided they are viewed as a function of the right variable. This is
not $q^2$ but $v{\cdot}v^{\prime}$ where $v$ is the 4-velocity ($p_{\mu}/m$)
of the initial meson and $v^{\prime}$ is the 4-velocity of the final
meson.  $v{\cdot}v^{\prime}$ is often given the symbol $w$. In the notation
of Equation~\ref{formfs} $w = (M_P^2 + M_{P^{\prime}}^2 -
q^2)/(2M_PM_{P^{\prime}})$.  The limit $w=1$ is known as the
`zero-recoil' limit because this corresponds to the kinematic point
where the $B$ meson at rest decays to, say, a $D$ meson at rest and
the decay products of the $W$ come out back-to-back.  This point has
maximum $q^2 = (M_P - M_{P^{\prime}})^2$.

\begin{figure}
\centerline{\includegraphics[clip,trim=0 50 0 60,width=148mm]{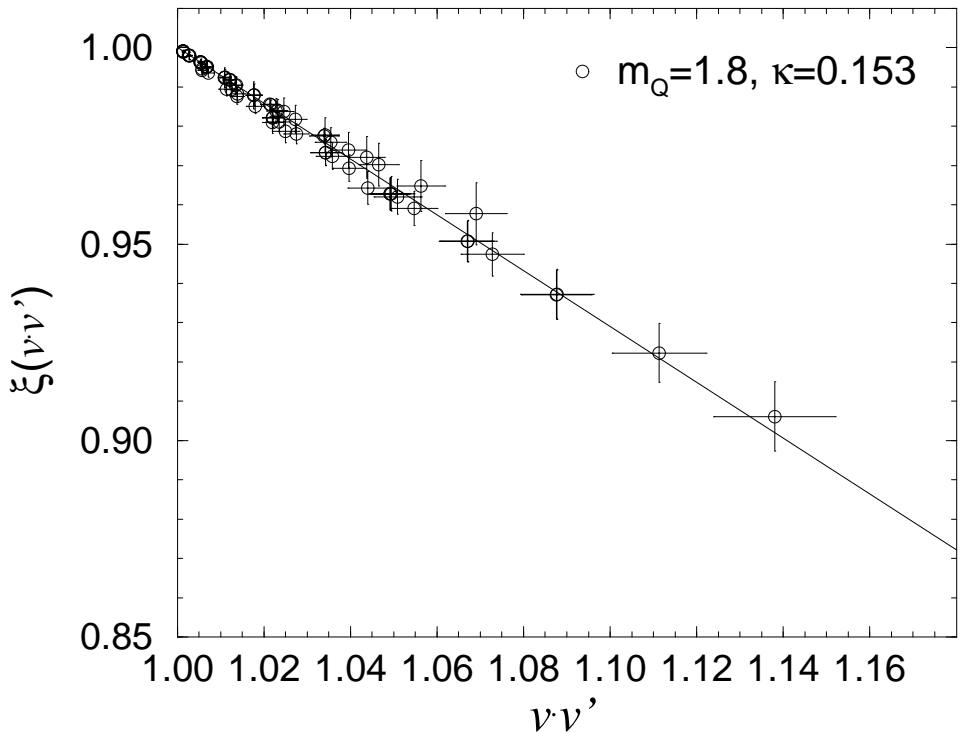}}
\caption{ The $B \rightarrow B$ elastic form factor calculated in
lattice QCD using NRQCD for the $b$ quark and plotted versus $w$
(Hashimoto, 1996). }
\label{hashi}
\end{figure}

The $B \rightarrow B$ elastic form factor takes the form
\begin{equation}
\langle B(v^{\prime}) | V_{\mu} | B(v) \rangle = M_B \xi(w) (v+v^{\prime})
\end{equation}
in the limit of infinite $b$ quark mass, where $\xi(w) = f_+(q^2)$,
$f_- = 0$.  $\xi(w)$ is known as the Isgur-Wise
function\index{Isgur-Wise function}.  $\xi(1) = 1$ is an absolute
normalisation in the continuum because $\overline{b}\gamma_{\mu}b$ is
a conserved current.  The lattice current is not a conserved one
(except for the NRQCD/static actions) but if we are interested only in
the shape of $\xi(w)$ we can renormalise to match 1 at $w=1$.  There
have been several calculations of the $B \rightarrow B$ form factor on
the lattice, for various heavy quark masses. \Figure{\ref{hashi}} shows
such a calculation using NRQCD with a mass close to that for the $b$
quark (Hashimoto, 1996).

\begin{figure}
\centerline{\includegraphics[height=7.0cm,clip,trim=0 10 0 20]{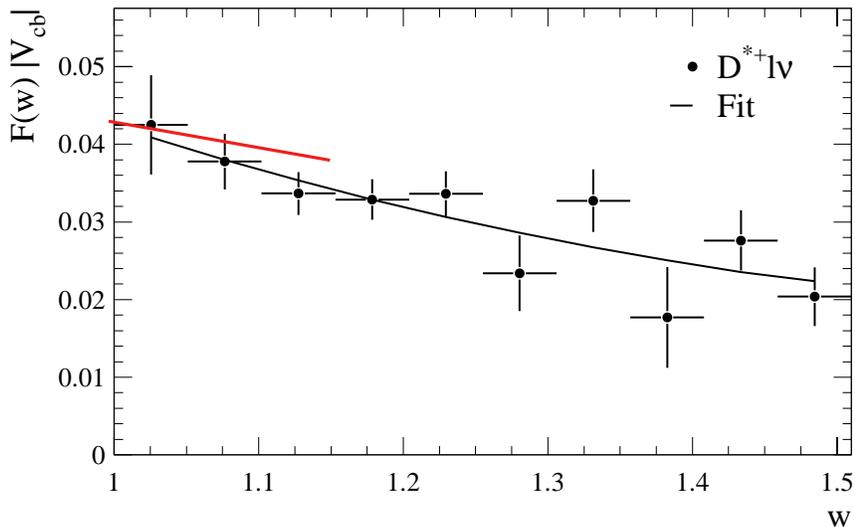}}
\caption{ $|V_{cb}|F(w)$ extracted from the experimental $B
\rightarrow D^{\myast}$ decay rate plotted as a function of $w$ (CLEO,
2000). The shorter curve on $1{<}w{<}1.15$ is a rescaled version of the curve in
Figure~\ref{hashi}. }
\label{cleo}
\end{figure}

The interest in calculating the Isgur-Wise function is that, in the
Heavy Quark Symmetry picture described above, it is also applicable to
$B \rightarrow D$ and $B \rightarrow D^{\myast}$ decays.  In these cases,
however, there is an additional overall perturbative renormalisation
because $\overline{b}\gamma_{\mu}c$ is not a conserved current, and
there are corrections which appear as differences of inverse powers of
the $b$ and $c$ quark masses.  For kinematic reasons, $B \rightarrow
D^{\myast}$ is experimentally easier to measure in the $w \rightarrow 1$
region. The differential decay rate is
\begin{equation}
\frac{d\Gamma}{dw} = |V_{cb}|^2 K(w){\cal F}^2(w)
\end{equation}
where $V_{cb}$ is the CKM element that we want to determine, $K(w)$ is
a kinematic factor and ${\cal F}$ is the form factor for the decay.
\Figure{\ref{cleo}} shows results from the CLEO collaboration (CLEO,
2000) for ${\cal F}$$(w)|V_{cb}|$.  The lighter hashed curve is the
result from the lattice shown in Figure~{\ref{hashi}} rescaled by a
constant to match at $w = 1$. Given lattice results for $B \rightarrow
D^{{\myast}}$ rather than $B \rightarrow B$, the constant required for
rescaling would be $|V_{cb}|$ which would then be determined.

In fact, a number of simplifications can be made to the lattice
calculation at the $w = 1$ point and so it is currently better to
perform a phenomenological extrapolation of the experimental data to
$w = 1$ and divide the extrapolated result by the lattice result for
$\cal F$$(1)$. The Fermilab group, using heavy relativistic (Fermilab)
quarks and ${O}$$(\alpha_s)$ matching to the continuum, give the
most precise result so far:
\begin{equation}
{\cal F}_{B \rightarrow D^{\myast}}(1) = 0.913{}^{+0.024}_{-0.017}{}^{+0.017}_{-0.030}
\end{equation}
in which the first error comes from statistics and fitting and the
second from systematic errors, including the effect of using the
quenched approximation (Hashimoto, 2001). The resulting value of
$|V_{cb}|$ extracted depends on which experiment's value for
$|V_{cb}|{\cal F}(1)$ is taken.  Using an average result (Stone, 2002)
of $37.8 \pm 1.4 \times 10^{-3}$ gives a value for $V_{cb}$ of $41.4
\pm 1.5 \pm 1.7 \times 10^{-3}$ where the final theoretical error
comes from adding the lattice errors in quadrature. The lattice and
experimental errors are currently of about the same size. The lattice
error can be improved further in an unquenched calculation with a
higher order matching of the lattice current to the continuum.

\begin{figure}[!h]
\centerline{\includegraphics[bb= 0 0 4196 3480, 
scale = 0.065,clip]{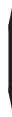}\rule{11mm}{0mm}}
\caption{\mbox{Lattice results for  the form factors for 
$B {\rightarrow }\pi$ decay (Bernard, 2001). }}
\label{formpi}
\end{figure}

$B \rightarrow$ light ($\pi, \rho$) semi-leptonic decay is rather
harder to calculate on the lattice. In many ways it is more important
to do, however, because continuum techniques, such as HQET, can give
very little useful input. One difficulty is that lattice systematic
errors are smallest where the $B$ and, say, the $\pi$ lattice momenta
are smallest, around the zero-recoil point discussed above, but there
is very little experimental data there. Most experimental data occurs
at relatively low $q^2$ values ($q^2 < 16 {\rm GeV}^2$) when the zero
recoil point has $q^2 = q^2_{\rm max} = (m_B - m_{\pi})^2$ = 26${\rm
GeV}^2$. A comparison of lattice results for the form factors for $B
\rightarrow \pi$ decay is shown in \Figure{\ref{formpi}} (Bernard,
2001).  Different lattice results are shown covering a range of
$q^2$. The reason that some results are at smaller $q^2$ than others
is because some use relativistic quarks (marked NPclover) at a mass
around the $c$ quark mass rather than the $b$.  For reasons discussed
earlier, none of the lattice calculations can be done at the physical
$u,d$ quark masses and so must be chirally extrapolated to that
point. This is done in a different way by different groups and has led
to very different final results, even though the intermediate data
does not show very different behaviour (see Figure~\ref{formpi}). A
better understanding of how the chiral extrapolation should be done
will be required before precise lattice results will be
available. Good experimental results in the $q^2$ region that the
lattice can reach will then allow a determination of $V_{ub}$.

\section{Conclusions}

Lattice QCD has come a long way from the original calculations of the
1970s. The original idea that we could solve a simple discretisation
of QCD numerically by `brute force' has been replaced by a more
sophisticated approach.  Unfortunately, to the uninitiated, this can
look like cookery.  I have tried to describe some of the calculational
and technical details so that non-practitioners feel able to make an
informed judgement about lattice calculations, and see where progress
will be made in the future. There is no doubt, for example, that
precise lattice calculations are needed to obtain maximum benefit from
the huge experimental investment in $B$ physics. In the next few years
such calculations will become possible, at least for some quantities,
and this will mark the `coming of age' of the lattice QCD approach at
last.

\section*{Acknowledgments} It was a pleasure to contribute to this lively
and interesting school.  I am grateful to all my collaborators, and
particularly Peter Lepage and Junko Shigemitsu, for numerous useful
discussions over many years. I am also grateful to Jack Cheyne, Greig
Cowan and Alan Gray for a critical reading of this manuscript.

\section*{References}
\frenchspacing
\begin{small}
\reference{Ali Khan A et al, 2002, CP-PACS collaboration, 
\textit{Light hadron spectroscopy with two flavors of dynamical quarks 
on the lattice}, Phys. Rev. D{\bf 65} 054505, hep-lat/0105015.}
\reference{Aoki S et al, 2000, CP-PACS collaboration, \textit{Quenched light hadron spectrum}, Phys. 
Rev. Lett. {\bf 84}, 238; hep-lat/9904012.}
\reference{Bernard C, 2001, \textit{Heavy quark physics on the lattice}, Nucl. Phys. B (Proc. Suppl. {\bf 94}) 159; hep-lat/0011064}
\reference{Bernard C et al, 2001, MILC collaboration, \textit{The QCD spectrum with three quark flavors}, 
Phys. Rev. D{\bf 64} 054506;
hep-lat/0104002.}
\reference{Buchalla G, 2002, \textit{Heavy quark theory}, this volume; hep-ph/0202106.}
\reference{Chen P, Liao X and Manke T, 2001, \textit{Relativistic quarkonia from anisotropic lattices},
Nucl. Phys. B (Proc. Suppl. {\bf 94}) 342; hep-lat/0010069.}
\reference{CLEO collaboration, 2000, \textit{Determination of the $B \rightarrow D^*l\nu$ decay width and $|V_{cb}|$}, hep-ex/ 0007052.}
\reference{Davies C, 1998, \textit{The heavy hadron spectrum} in \textit{Computing Particle Properties}, Springer LectureNotes in Physics, Eds Gausterer, Lang; hep-ph/9710394.}
\reference{Di Pierro M, 2001, \textit{From Monte Carlo Integration to Lattice Quantum Chromodynamics}, hep-lat/0009001}
\reference{Eichten E and Hill B, \textit{An effective field theory for the 
calculation of matrix elements involving heavy quarks}, Phys. Lett. B{\bf 243}, 511.}
\reference{El-Khadra A, Kronfeld A and Mackenzie P, 1997, \textit{Massive fermions
in lattice gauge theory}, Phys. Rev. D{\bf 55}, 3933; hep-lat/9604004.}
\reference{Gupta R, 1998, \textit{Introduction to Lattice QCD}, hep-lat/9807028.}
\reference{Hashimoto S and Matsufuru H, 1996, \textit{Lattice heavy quark effective theory and the Isgur-Wise function}, Phys. Rev D{\bf 54} 4578; hep-lat/9511027.}
\reference{Hashimoto S et al, 2001, \textit{Lattice calculation of the 
zero-recoil form factor of $B \rightarrow D^*l\nu$: towards a model
independent determination of $|V_{cb}|$}, hep-lat/0110253.}
\reference{Hein J et al, 2000, \textit{Scaling of the $B$ and $D$ meson spectrum
in lattice QCD}, Phys. Rev. D{\bf 62} 074503; hep-ph/0003130.}
\reference{Hocker A, 2001, \textit{A new approach to gloabl fit of the 
CKM matrix}, Eur. J. Phys. C{\bf 21} 25; hep-ph/0104062; http://www.slac.stanford.edu/\~{}laplace/ckmfitter.html.}
\reference{LAT2000, Nucl. Phys. B (Proc. Suppl. {\bf 94}) 2001.}
\reference{LAT2001, Nucl. Phys. B (Proc. Suppl. {\bf 106}) 2002.}
\reference{Lepage P and Mackenzie P, 1993, \textit{On the viability of 
lattice perturbation theory}, Phys. Rev. D{\bf 48}, 2250; hep-lat/9209022.}
\reference{Lepage P et al, 1992, \textit{Improved nonrelativistic QCD for 
heavy quark physics}, Phys. Rev. D{\bf46}, 4052; hep-lat/9205007.}
\reference{Maynard C et al, 2002, UKQCD collaboration, \textit{Heavy-light 
decay constants on the lattice}, Nucl. Phys. B (Proc. Suppl. {\bf 106}), 388; hep-lat/0109026.}
\reference{Lubicz V, 2001, \textit{Quark masses on the lattice, light and heavy},
Nucl. Phys. B (Proc. Suppl. {\bf 94}) 116; hep-lat/0012003.}
\reference{Marcantonio L et al, 2001, UKQCD collaboration, \textit{The unquenched
$\Upsilon$ spectrum}, Nucl. Phys. B (Proc. Suppl. {\bf 94}), 363; hep-lat/0011053.}
\reference{PDG, 2001, http://pdg.lbl.gov/.}
\reference{Rosner J, 2002, \textit{The Standard Model in 2001}, this volume; hep-ph/0108195.}
\reference{Ryan S, 2002, \textit{Heavy quark physics from lattice QCD}, Nucl. Phys. B (Proc. Suppl. {\bf 106}) 86; hep-lat/0111010.}
\reference{Sommer R, 1998, \textit{Non-perturbative renormalisation of QCD} in \textit{Computing Particle Properties}, Springer Lecture Notes in Physics, Eds Gausterer, Lang; hep-ph/9711243.}
\reference{Stone S, 2002, \textit{B Phenomenology}, this volume; hep-ph/0112008.}
\reference{Toussaint D, 2002, \textit{Spectrum results with Kogut-Susskind 
quarks}, Nucl. Phys. B (Proc. Suppl. {\bf 106}) 111; hep-lat/0110010.}
\end{small}
\end{document}